\documentclass[11pt]{article}
\pdfoutput=1
\usepackage{jheppub}
\usepackage{float, extarrows, tikz-cd}
\usepackage{graphicx}
\usepackage{slashed}
\usepackage{tabularx,ragged2e}
\usepackage{amssymb}
\usepackage{subcaption}
\usepackage{amsmath,amssymb}
\usepackage{slashed}
\usepackage{caption}
\usepackage{xcolor}
\usepackage{dsfont}
\usepackage{verbatim}
\usepackage{mathtools, xcolor,ytableau, amsfonts,tikz}
\usepackage{graphicx}
\usepackage{physics}
\newcolumntype{C}{>{\Centering\arraybackslash}X}

\numberwithin{equation}{section}

\def\le{\left}
\def\ri{\right}
\newcommand\ov{\over}
\newcommand\p{\ensuremath{\partial}}
\newcommand{\es}[2] {\begin{equation} \label{#1} \begin{split} #2 \end{split} \end{equation}}

\def\<{\langle}
\def\>{\rangle}

\newcommand\om{\omega}
\newcommand\Om{\Omega}
\newcommand\ga{{\ensuremath{{\gamma}}}}
\newcommand\Ga{{\ensuremath{{\Gamma}}}}
\newcommand\de{{\ensuremath{{\delta}}}}
\newcommand\De{{\ensuremath{{\Delta}}}}
\newcommand\sO{{\ensuremath{{\mathcal O}}}}
\newcommand\sM{{\ensuremath{{\mathcal M}}}}
\newcommand\Vol{{\ensuremath{{\mathrm{Vol}}}}}
\newcommand {\be} {\begin {equation}}
\newcommand {\ee} {\end {equation}}

\ytableausetup{centertableaux,boxsize=.5em}

\title{The volume of the black hole interior at late times}

\author{Luca V. Iliesiu${}^1$, M\'ark Mezei$^2$ and G\'abor S\'arosi${}^3$ }

\affiliation[1]{Stanford Institute for Theoretical Physics, Stanford University, Stanford, CA 94305, USA}
\affiliation[2]{Simons Center for Geometry and Physics, SUNY, Stony Brook, NY 11794, USA}
\affiliation[3]{CERN, Theoretical Physics Department, 1211 Geneva 23, Switzerland}

\abstract{
Understanding the fate of semi-classical black hole solutions at very late times is one of the most important open questions in quantum gravity. In this paper, 
we provide a path integral definition of the volume of the black hole interior and study it 
at arbitrarily late times for black holes in various models of two-dimensional gravity. Because of a novel universal cancellation between the contributions of the semi-classical black hole spectrum and some of its non-perturbative corrections, we find that, after a linear growth at early times, the length of the interior saturates at a time, and towards a value, that is exponentially large in the entropy of the black hole. This provides a non-perturbative confirmation of the complexity equals volume proposal since complexity is also expected to plateau at the same value and at the same time. 
}
\begin{document}
\begin{flushright}
\hfill{\tt CERN-TH-2021-102}
\end{flushright}

\maketitle
\renewcommand{\arraystretch}{1.5}
\newcommand{\cev}[1]{\reflectbox{\ensuremath{\vec{\reflectbox{\ensuremath{#1}}}}}}
\newcommand{\lag}{\mathcal{L}}
\newcommand{\beq}{\begin{equation}\begin{aligned}}
\newcommand{\eeq}{\end{aligned}\end{equation}}
\newcommand{\ti}{\frac{2i}{3}}
\newcommand{\tp}{\frac{1}{2\pi}}
\newcommand{\tb}{\tilde{b}}
\newcommand{\fp}{\frac{1}{4\pi}}
\newcommand{\dc}{\Delta c}
\newcommand{\lz}{\mathcal{L}_0[A]}
\newcommand{\cT}{\mathcal T}
\newcommand{\lh}{\hat{\mathcal{L}}}
\newcommand{\cM}{{\mathcal{M}}}
\newcommand{\nsi}{\text{non-int.}}
\newcommand{\gam}{\gamma}
\newcommand{\cG}{\mathcal G}
\newcommand{\Gam}{\Gamma}
\newcommand{\Mod}{\text{Mod}}
\newcommand{\cW}{\mathcal W}
\newcommand{\til}{\tilde}
\newcommand{\pp}[2]{\frac{#1}{\sqrt{1+(#2)^2}}}
\newcommand{\bpm}{\begin{pmatrix}}
\newcommand{\epm}{\end{pmatrix}}
\newcommand{\texr}{\textcolor{red}}
\newcommand{\cC}{\mathcal{C}}
\newcommand{\lr}{\left\langle}
\newcommand{\rr}{\right\rangle}

\section{Introduction}

Among the greatest challenges in physics is obtaining a microscopic understanding of the expanding universe. While this is presently out of reach for the observed universe, an intriguing example of such an expanding spacetime is the interior of black holes, which in turn can be studied in the framework of the AdS/CFT correspondence \cite{Maldacena:1997re,Witten:1998qj,Gubser:1998bc}.  It has been proposed by Susskind and collaborators that the growth of the black hole interior is dual to the growth of complexity, the minimal number of gates required for a quantum circuit to prepare the corresponding time evolved  state in the dual CFT \cite{Susskind:2014rva,Susskind:2014moa}. 
One version of this proposal is that complexity equals the volume of a maximal slice in the black hole interior \cite{Stanford:2014jda}.\footnote{Both the definition of complexity and the size of the black hole interior suffer from many ambiguities, which have lead to many versions of this proposal \cite{Brown:2015bva,Brown:2015lvg,Couch:2016exn,Belin:2021}. However, these ambiguities are believed not to affect the late time universal scaling properties. } The most important rationale behind this conjecture is that the black hole interior grows for a very long time, past when conventional probes, like correlation functions or entanglement entropy, have already reached their thermal equilibrium value. Complexity, on the other hand, is expected to grow in chaotic systems at the same rate as the volume of the interior, until times exponentially large in the entropy, after which it saturates in a complexity plateau \cite{Susskind:2015toa,Brown:2016wib,Balasubramanian:2019wgd,Susskind:2020wwe,Balasubramanian:2021mxo, Haferkamp:2021uxo}. Up to this point, it has been an open problem to establish whether the volume of the black hole interior exhibits a similar saturation at late times. 

A priori, there is no reason to expect that a geometrical description of the interior could capture the late time saturation of complexity. However, recent developments show that the gravitational path integral has a surprisingly large regime of validity, with the exponentially small corrections coming from summing over topologies being capable of capturing fine details of spectral statistics \cite{Saad:2019lba,Cotler:2020ugk,Belin:2020hea}, or the unitary Page curve of the entropy of an evaporating black hole \cite{Almheiri:2019qdq,Penington:2019kki}.  Emboldened by this success, in this paper, we calculate the corrections coming from higher topologies to the volume of the black hole interior in a wide class of models of two-dimensional gravity, including the widely studied theory of JT gravity \cite{Teitelboim:1983ux,Jackiw:1984je,Maldacena:2016upp}. Importantly, this requires us to give a nonperturbative definition of the length of the black hole interior and to complete the sum over surfaces of higher topology to the matrix model of Saad-Shenker-Stanford \cite{Saad:2019lba} (or its generalizations \cite{Witten:2020wvy, Maxfield:2020ale}). 
We indeed find that the length of the black hole interior saturates at a time and length exponentially large in the entropy. Our result thus provides a non-perturbative version of the complexity equals volume conjecture.

In section \ref{sec:complexity-in-dilaton-grav} we propose a natural candidate for the length $\< \ell\>$ of the Einstein-Rosen (ER) bridge. In a two-dimensional theory, the volume of the black hole interior is given by length of the geodesic going from one side to the other. While the definition of this length is unambiguous on surfaces with trivial topology, on manifolds of higher genus there are an infinite number of geodesic spanning between the two sides of the ER bridge. We explore various definitions for this length and settle on one that is well defined on surfaces of arbitrary topology, causes minimal backreaction on the metric, and can be analytically continued between Euclidean and Lorentzian geometries:
\es{regsumIntro}{
\<\ell\>=\lim_{\De\to0}\expval{\sum_{\ga} \ell_\ga \, e^{-\De \ell_\ga}}\,,
} 
where $\ga$ is labeling non self-intersecting geodesics, $\<\dots\>$ evaluates the gravitational path integral by summing over surfaces with any topology 
and where $\De$ acts as a regulator. Since the free field boundary-to-boundary propagator on hyperbolic surfaces is given by the sum over geodesics (including self-intersecting ones), $\<\ell\>$ is closely related to the two-point function $\<\chi \chi\>$, with an operator of dimension $\De$ inserted on each side of the two-sided black hole, as a probe for the black hole interior.

In section \ref{sec:two-point-JT}, we turn to the analysis of this quantity.
In the limit in which $\chi$ is a probe field and when we approximate the two-point function by summing over non self-intersecting geodesics, 
 we prove that at arbitrary  order in $e^{S_0}$ (with $S_0$ being the leading order entropy of the black hole) the two-point function is simply given by 
\es{intro-two-point-formula}{
\< \chi(x_1) \chi(x_2) \>_\nsi& \sim \int_0^\infty dE_1 dE_2 \ \le\<\rho(E_1)\rho(E_2) \ri\> e^{-E_1 (x_1-x_2)- E_2(\beta-x_1+x_2)} \, \sM_\De(E_1,E_2)\,,
}
where $\sM_\De(E_1,E_2)=\abs{\<E_1|\chi|E_2\>}^2$ can be computed analytically. This result can be viewed as a generalization to arbitrary genus and dilaton potential of the genus 1 result for JT gravity of \cite{Saad:2019lba}. 
 The expression \eqref{intro-two-point-formula} is also consistent with a conjectured formula for the matter two-point function in gravity that was obtained by studying operator insertions in the matrix integral that is dual to the JT theory  \cite{Saad:2019lba, Blommaert:2020seb}.

 \begin{figure}[!t]
\centering
\includegraphics[width=0.65\textwidth]{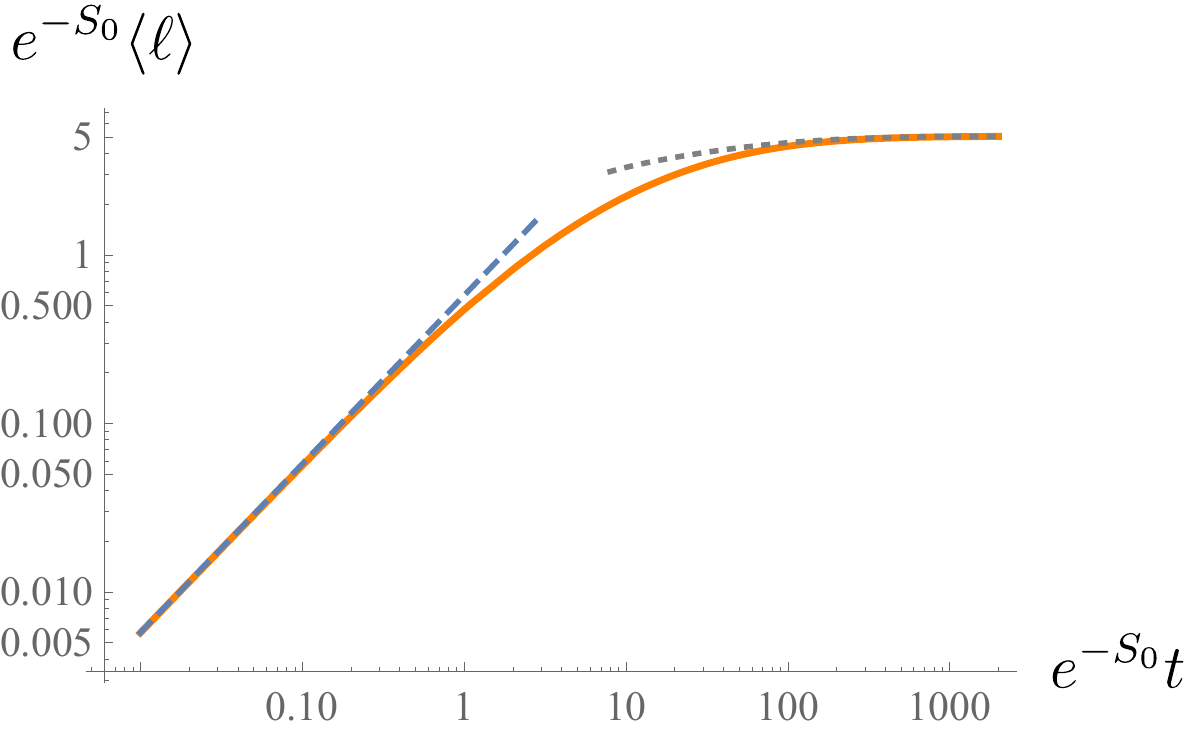} 
\caption{We show the numerical evaluation of the Einstein-Rosen bridge length (see \eqref{Ct3}) for $\beta=15$ in orange, while the early and late time asymptotics \eqref{Asymp-ell} are shown by the blue dashed and gray dotted curves, respectively.}
\label{fig:1}
\end{figure} 
 
In section \ref{sec:evaluation} we obtain the main result of our paper, that is, up to an overall additive constant\footnote{The linear growth of the black hole interior at time scales $t\ll e^{S_0}$ but exactly in the temperature was first obtained in \cite{Yang:2018gdb}. }
\es{Asymp-ell}{
\<\ell(t)\> \approx 
\begin{cases}
C_1 t+\dots \qquad&  t\ll e^{S_0}\,,\\
C_0-\dots & t\gg e^{S_0}\,,
\end{cases}
}
in a large class of theories of dilaton gravity, including JT gravity. Here, the coefficients $C_1$ is $O(1)$ while $C_0$ is $O\le(e^{S_0}\ri)$ and depend on the spectral density $\rho(E)$ in each dilaton gravity model and on the inverse-temperature $\beta$ of the thermofield double state in which the system is prepared at $t=0$. For a plot of $\<\ell(t)\>$ see figure~\ref{fig:1}.

In section  \ref{sec:comparison-section} we remark on the late-time differences between the ER bridge length, the spectral form factor, and the two-point function. While all three quantities exhibit a period of growth and then a plateau at later times, the physics of the growth and plateau periods for the bridge is widely different from that for the latter two quantities. While the period of growth for the ER bridge length is due to the contribution from the classical solution, the ``ramp'' in the  spectral form factor and two-point function is due to the leading Euclidean wormhole geometry.  More importantly, while the ``plateau'' in the spectral form factor and two-point function occurs due to a cancellation between the  wormhole geometry and other non-perturbative corrections to the spectral two-point function $\le\<\rho(E_1)\rho(E_2) \ri\> $, the saturation of the bridge length is due to novel cancellations between the classical disconnected contribution and non-perturbative contributions in $\le\<\rho(E_1)\rho(E_2) \ri\> $.  To our knowledge, these types of cancellations have not been noticed in the random matrix literature before and are universal in all one matrix models.

To further probe the relation between complexity and the late time length of the ER bridge, we compute the variance of the latter quantity  in section \ref{sec:variance-ER-bridge}. We start by deriving a geometric formula for the variance in terms of the spectral four-point function. We then show that at late times 
\es{var-overall-result}{
\sigma_{ \ell}\sim e^{S_0/2}\,, \qquad \text{for}\quad  t\sim e^{S_0}\,, \qquad\,\,\, \text{ and }\,\,\, \qquad \sigma_{ \ell}\sim \sqrt{t}\,, \qquad \text{for}\quad  t\gg e^{S_0}\,,
}
In contrast to the plateaus for the spectral form factor and the two-point function, on the plateau for the ER bridge, the ``signal'' is much larger than the ``noise'' in the regime $t \ll e^{2S_0}$. However, we find that as opposed to $\langle \ell \rangle$, the variance $\sigma_\ell$ does not have a plateau, instead it grows forever, in particular, it becomes the same order as the signal when $t\propto e^{2S_0}$. We will discuss some possible explanations for this unexpected behavior.

Finally, in section \ref{sec:discussion} we discuss the interpretation of our results beyond the theories of dilaton gravity and their matrix integral dual. Our geometric computation suggests the importance of a new spectral quantity, called spectral complexity, which we expect to closely track the volume of the black interior in any chaotic system with a holographic dual. We confirm this expectation in a numerical study of the SYK model.

\section{Nonperturbative length}
\label{sec:complexity-in-dilaton-grav}

We want to define the length $\expval{\ell}$ of the ER bridge in two-dimensional models of gravity. Our definition needs to take into account that in higher genus geometries, there are multiple geodesics connecting two boundary points. Furthermore, when we probe this length we do not wish to experience strong back-reaction from the measurement since that might spoil the growth of the ER bridge.

 One is first tempted to consider the minimal length geodesic spanning between the two sides of the ER bridge as the definition of $\ell$. However, in a gravitational path integral, we oftentimes have to work with complexified geometries and complex geodesic lengths. Consequently, the minimum geodesic length cannot be analytically continued.  Alternatively, one might imagine computing the minimal length of the ER bridge by considering the two-sided two point function $\< \chi_L(t) \chi_R(0) \>$ and taking the limit
 $-\lim_{\Delta \to \infty} \frac{1}{\Delta} \log \< \chi_L(t) \chi_R(0) \>= \lim_{\Delta \to \infty}   \frac{1}{\Delta} \<\sum_{\ga} \, e^{-\De \ell_\ga}\>$ which seemingly picks the geodesic with minimal real part from the sum.\footnote{Here, $L/R$ denotes operators inserted on the left/right sides of the bridge. } However, due to the backreaction of the operator insertion on the metric the measurement of the ER bridge length is spoiled:  as shown in \cite{Yang:2018gdb} for any $\Delta>0$, the two-point function decays exponentially up to a time $t \sim O(1)$, past which  it transitions to a power law decay due to quantum corrections coming from JT gravity ($\sim 1/t^3$, regardless of the value of $\Delta$). Because of this, $-\lim_{\Delta \to \infty} \frac{1}{\Delta} \log \< \chi_L(t) \chi_R(0) \>$ does not see the growth of the ER bridge past the thermalization time (which is $\sim \beta$). Therefore, we dismiss these candidates when defining the length of the bridge.  
 
A  candidate that resolves all the issues above is  the regularized sum\footnote{We will specify the exact meaning of $\<\dots\>$ shortly. }
\es{regsum}{
\<\ell\>=\lim_{\De\to0}\expval{\sum_{\ga} \ell_\ga \, e^{-\De \ell_\ga}}\,,
} 
where $\ga$ is labeling non self-intersecting geodesics and $\<\dots\>$ evaluates the sum over geodesics within the JT gravity or dilaton gravity path integral, including the contribution of surfaces of arbitrary genus. The prescription for choosing geodesics without self-intersections is natural since, if we cut the manifold open along such curves, we obtain a state in which the boundary points are separated by geodesic length $\ell_\ga$.\footnote{The cut that separates the manifold into two disjoint parts will in general also create some accompanying baby universes.} From the quantum information perspective, this choice makes contact with the original motivation of interpreting $\expval{\ell}$ as a length of a minimal tensor network that can create the state. 
Furthermore, this quantity can be analytically continued and has minimal backreaction on the metric.\footnote{We can think about this quantity as inserting a tensionful defect and extracting the leading correction in the tensionless limit. This is similar to how the minimal surface is obtained from the gravitational R\'enyi entropies using the replica trick \cite{Lewkowycz:2013nqa}.} However, one might worry that \eqref{regsum} could be divergent since it involves an infinite number of geodesics on all surfaces with $g>1$. As we show shortly, we find that the time-dependent piece in \eqref{regsum} is finite, independent of the regularization scheme, and is precisely the probe of the late time physics that we are interested in:  i.e., it is sufficient to study $\< \ell(t)\> - \<\ell(0)\>$.

We will compute $\<\ell(t)\>$ by taking the $\De$ derivative of the two-sided correlation function, computed in an approximation, where we drop self-intersecting geodesics. Explicitly, 
\es{regsum3}{
\<\ell(t) \>= -\lim_{\De\to0}{ \frac{ \partial \le\< \chi_L(t) \chi_R(0)\ri\>_\nsi }{\p \De}}\,.
} 
Two-sided correlation functions can be obtained by analytic continuation from Euclidean correlators, and it will be convenient to obtain $\le\< \chi_L\le(t\ri) \chi_R\le(0 \ri) \ri\>$ from a Euclidean computation performed in section \ref{sec:two-point-JT}. The analytic continuation that we use is valid in any quantum system
\es{AnalyticCont}{
\le\< \sO_L(t) \sO_R(0) \ri\>&\equiv \bra{\text{TFD}_\beta} \sO_L(t) \sO_R(0) \ket{\text{TFD}_\beta}=\bra{\text{TFD}_\beta} \sO_L(t/2) \sO_R(-t/2) \ket{\text{TFD}_\beta}\\
&=\Tr\le[\sO\le({\beta/ 2}+i t\ri)\sO(0) e^{-\beta H}\ri]\equiv \le\< \sO\le({\beta/ 2}+i t\ri) \sO(0) \ri\>_\beta\,.
}
where in the first line we used the two-sided (thermofield double) and in the second the Euclidean representation. The last expression in the first line is the most intuitively connected to the ER bridge length where we evolved both the left and rightt boundary points forward in time equally by $t/2$.\footnote{Recall that in the thermofield double in the right system the natural time direction runs backwards.} The equality between the first and second lines can be established by inserting a complete set of states in the second line and using that the thermofield double state is $ \ket{\text{TFD}_\beta}=\sum_n e^{-\beta R_n/2}\ket{E_n}_L\ket{E_n^*}_R$.  The dilaton gravity theories considered in this paper are famously ensembles of quantum systems, random matrix theories, so when we apply \eqref{AnalyticCont} in our context, the gravity computation gives the ensemble average of the above quantities. 

In section \ref{sec:two-point-JT}, we will shown that $\le\< \chi(x_1) \chi(x_2) \ri\>_\nsi$ is a particularly nice quantity in JT gravity and other theories of dilaton gravity, as it can be computed to all orders in the genus expansion and can even be computed non-perturbatively using the matrix model representation for such gravitational theories. We will therefore use  \eqref{regsum3} to extract $\<\ell (t)\> - \<\ell (0)\>$.

\section{The two-point function in JT gravity}
\label{sec:two-point-JT}

JT gravity is a theory of two-dimensional gravity with a metric $g_{\mu \nu}$, coupled to a dilaton $\phi$, whose Euclidean action is given by \cite{Teitelboim:1983ux,Jackiw:1984je,Maldacena:2016upp}
\be 
\label{eq:JT-action}
I_{JT} = -S_0 \chi(\cM) - \frac{1}2 \int \sqrt{g} \phi(R+2) - \int_{\partial \cM} \sqrt{h} \phi(K-1)\,, \qquad Z = \int \frac{Dg_{\mu \nu} D\phi}{\text{Diffeos}} e^{-I_{JT}}.
\ee
The first term is purely topological, as it only accounts for the Euler characteristic of the manifold $\cM$. The bulk term fixes the geometry to be a patch of hyperbolic space whose boundary has fixed, large, proper length $L = \beta/\epsilon$ and dilaton value $\phi= 1/\epsilon$.\footnote{Since we set the boundary value of the dilaton to one, times and inverse temperatures in the rest of the paper are measured in these units. In particular the Schwarzian theory is strongly coupled when $\beta\gg 1$.} These boundary conditions require the presence of the Gibbons-Hawking-York term, which gives an action for the boundary fluctuations in each such hyperbolic patch. While the model exhibits great simplicity since it only depends on topological data and boundary fluctuations, it still has black hole solutions and, in particular, we will be interested in studying two-sided black holes, which have an ER bridge between the two sides.  

In this section, our goal is to probe the length of the ER bridge when one includes the contribution of surfaces with arbitrary topology to the gravitational path integral in \eqref{eq:JT-action}.  The probe that we shall consider is the boundary-to-boundary two-point function of a probe matter field $\chi$, with scaling dimension $\Delta$, which only couples to the metric in the JT theory.  In the probe limit,   the two-point function on any hyperbolic surface can be schematically expressed as
\be 
\<\Tr_{\beta}\le(\chi(x_1) \chi(x_2)\ri)\> =\sum_{\mathcal M}  \sum_{\gamma} \<\Tr_\beta(e^{-\Delta \ell_{\gamma}})\>_{\mathcal M}\,,
\ee 
where the sum over $\cM$ is a sum over all surfaces, obeying the boundary conditions discussed above, and the sum over $\gamma$ is a sum over all geodesics connecting the points at boundary times $x_1$ and $x_2$, each of length $\ell_\gamma$. 

To clarify our notation for expectation values of observables in the gravitational path integral, $\Tr_\beta(\dots) \equiv \Tr(e^{-\beta H} \dots)$ is a mnemonic for the insertion of a gravitational boundary with the boundary conditions described above while the $\dots$ will indicate what operators are inserted on each boundary. Finally, the $\<\dots\>$ will represent the evaluation of the gravitational path integral with a number of boundaries given by the number of traces inside, each with its own specified operator insertions.

 On a hyperbolic surface of genus $g\geq 1$, there will be an infinite number of geodesics $\gamma$, some of which will self-intersect. Since we are interested in the two-point function in order to probe the length $\ell_\gamma$ of the ER bridge in a two-sided black hole, we shall restrict ourselves to geodesics that do not self-intersect and set $x_1 =0$ and $x_2 = \frac{\beta}2 +it$; for each surface $\cM$ we will denote such geodesics going from $x_1$ to $x_2$ by $\cG_{x_1, x_2}$ and will denote correlators that solely account for these geodesics by $\< \dots\>_\nsi$.\footnote{Since we discarded self-intersecting geodesics, $\<\Tr_\beta\le(\chi(x_1) \chi(x_2)\ri)\>_\nsi$ is only an approximation to the full two-point function. It was argued by Saad \cite{Saad:2019pqd} that this approximation gives the leading behavior of the function for large Lorentzian times, and he computed this modified two-point function to the first subleading order in $e^{S_0}$ given by the genus 1 geometry. The leading disk contribution (which does not suffer from this ambiguity) was obtained in \cite{Mertens:2017mtv}. Since in this paper, our focus is on studying the length of the ER bridge and not on the full two-point function, we will not further discuss whether self-intersecting geodesics are important in its late-time behavior.  }  
 
\begin{figure}[t!]
\centering
\includegraphics[width=0.9\textwidth]{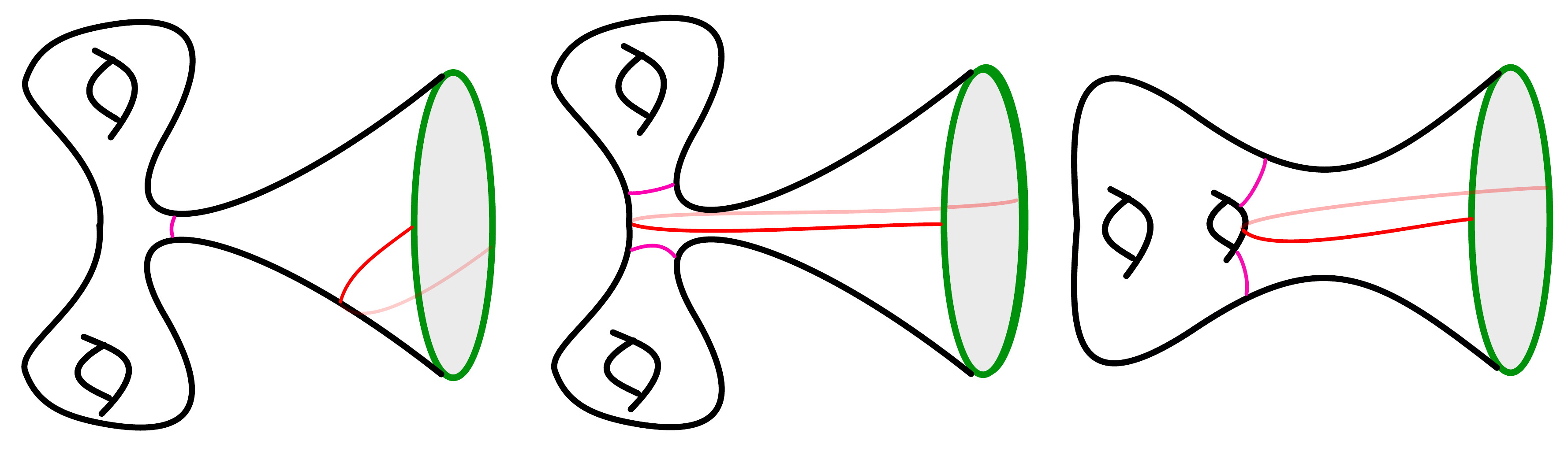} 
\caption{An example (with genus 2) of the type of hyperbolic surfaces which the path integral \eqref{general-formula} sums over. The green curves represent the wiggly asymptotic boundaries, while the red curve represents the boundary to boundary geodesic. While there are an infinite number of non-self-intersecting boundary-to-boundary geodesics on such surfaces, above, we have drawn an example of all possible surface topologies that can result by cutting a genus two surface along the geodesics. The purple curves represent the closed geodesics which we use to glue the trumpet wavefunctions in \eqref{two-point-function-vol-decomp} to different bordered Riemann surfaces.   }
\label{fig:example-at-genus-2}
\end{figure}
 
 In JT gravity, the path integral over all metrics reduces to an integral over boundary wiggles which we denote by $D{(\cW)}$ and an integral over the moduli space of hyperbolic surfaces \cite{Saad:2019lba}. Therefore, the two-point function can be rewritten as
\es{general-formula}{
\< \Tr_\beta\le(\chi(x_1) \chi(x_2) \ri)\>_{\nsi} =  \sum_g e^{S_0(1-2g)} \int_{\frac{\cT_{g,1}}{\Mod(\cM_{g,1})}} \omega \int D{(\cW)} e^{-I_{JT,\text{ bdy}}(\cW)}  \sum_{\gamma \in \cG_{x1, x2}} e^{-\Delta \ell_{\gamma}}\,,
 }
 where $\omega = \sum_{j=1}^{3g-3+n}db\wedge  d\tau$ is the Weil-Peterson symplectic form  on the moduli space of hyperbolic surfaces, and  $\cT_{g,1}$ represents the moduli space of all hyperbolic surfaces $\cM_{g,1}$ of genus $g$ with $1$ asymptotic boundary. Due to the redundancy of large diffeomorphism  in the gravitational path integral, we are supposed to quotient this moduli space by the mapping class group $\Mod(\cM_{g,1})$ of $\cM_{g,1}$. As in the example shown in figure \ref{fig:example-at-genus-2}, cutting along each geodesic $\gamma$ in the sum above, we can either obtain two disconnected manifolds, each with a single boundary, $\cM_{h,1}$ and $\cM_{g-h, 1}$ with $0\leq h \leq g$, or a single connected manifold with two boundaries $\cM_{g-1,\, 2}$.  The mapping class group acts by relating all boundary-to-boundary geodesics within a given cutting class as well as all closed geodesics on $\cM_{h,1}$ and $\cM_{g-h, 1}$ (or $\cM_{g-1,\, 2}$ in the latter case) which do not intersect the boundary-to-boundary geodesic. Therefore, following a similar strategy as in \cite{mirzakhani2007simple, Saad:2019pqd, Blommaert:2020seb}, we can rewrite the integral over the moduli space as 
 \es{rewriting-MCG}{
 \int_{\frac{\cT_{g,1}}{\Mod(\cM_{g,1})}} \omega \sum_{\gamma \in \cG_{x1, x2}} e^{-\Delta \ell_{\gamma}} =e^{-\Delta \ell} \int_{\frac{\cT_{g-1,2}}{\Mod(\cM_{g-1,2})}} \omega + \sum_{h\geq 0} e^{-\Delta \ell}\int_{\frac{\cT_{h,1}}{\Mod(\cM_{h,1})}} \omega \int_{\frac{\cT_{g-h,1}}{\Mod(\cM_{g-h,1})}} \omega\,,
 }
 where $\ell$ is the length of a geodesic which is now part of the boundary of  $\cM_{h,1}$ and $\cM_{g-h,1}$ (or part of both boundaries of $\cM_{g-1,2}$) and where for the remainder of the paper the sum over $h$ is bounded such that disconnected surfaces of genus $h$ and $g-h$ are not counted twice. When the genus $g= 2$, we can use figure \ref{fig:example-at-genus-2} to better understand \eqref{rewriting-MCG}.  The first term on the RHS sums over the moduli space of the genus one surface with two boundaries obtained after cutting the rightmost figure along the red geodesic. The $h=0$ term in \eqref{rewriting-MCG} corresponds to the sum over the moduli space of the leftmost figure after a cut along the red geodesic which results into a one boundary genus zero and a one boundary genus one surface, while the $h=1$ term correspond to summing over the moduli space of the two genus one surfaces obtained after the cut in the center figure.

 As drawn in figure \ref{fig:example-at-genus-2},  we can now cut each such surface along the closed geodesic which is closest to the asymptotic boundary to separate the manifolds, into two bordered Riemann surfaces with a geodesic boundary (or a single such surface in the case of $\cM_{g-1,2}$), and two trumpets which, on one end, have a closed geodesic boundary, and, on the other end, have a boundary with fixed proper length and fixed dilaton glued to a geodesic of fixed length, $\ell$. Following the wavefunctional formalism discussed in \cite{Yang:2018gdb},  we denote the path integral for such an object  as $\psi_{\text{Trumpet},x}(\ell, b)$. The expression for the trumpet wavefunction was derived in \cite{Saad:2019pqd} in terms of the Hartle-Hawking wavefunction for a disk with a fixed ADM mass $E$, cut by a geodesic boundary of length $\ell$, $\psi_{\text{Disk, E}}(\ell)$:
\es{wavefunction-trumpet}{
\psi_{\text{Trumpet, } x}(\ell, b)  = \begin{tikzpicture}[baseline={([yshift=-.0ex]current bounding box.center)}, scale=0.6]
 \pgftext{\includegraphics[scale=0.5]{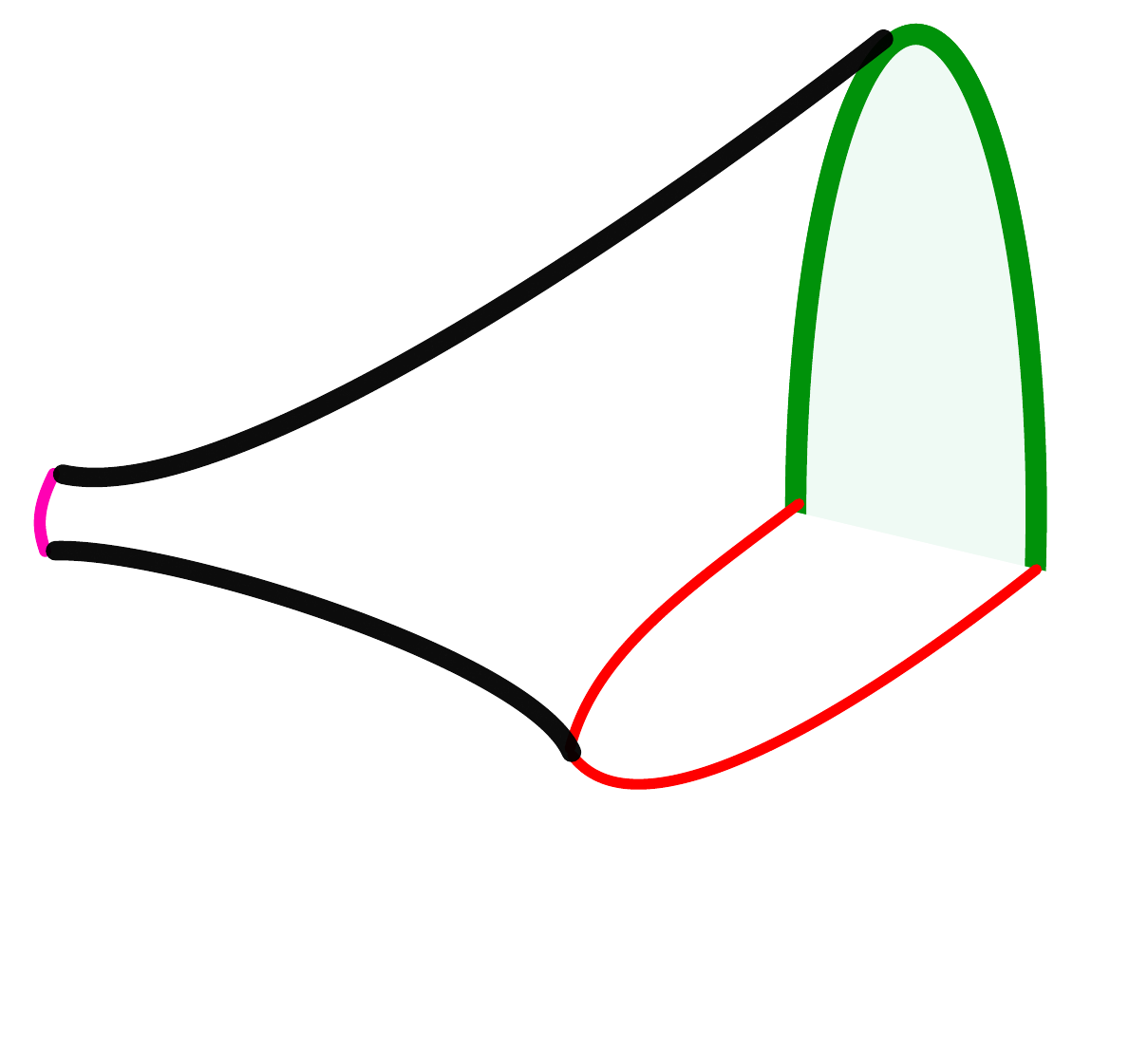}} at (0,0);
   \draw (-3.35,0) node  {$b$}; 
    \draw (0.5,-1.7) node  {$\ell$}; 
    \draw (2.8,1.7) node  {$x$}; 
  \end{tikzpicture}
  = \int_0^\infty dE \rho_{\text{Trumpet}}(E) \psi_{\text{Disk}, \,E}(\ell) e^{-x E} \,,
}
where the density of states of the trumpet  $\rho_{\text{Trumpet}}(E)= \frac{\cos(b\sqrt{2E})}{\pi \sqrt{2E}}$  can be computed from fermionic localization \cite{Stanford:2017thb}, 
while $\psi_{\text{Disk}, E}(\ell)$ is given by
$
\psi_{\text{Disk}, E}(\ell) = 4 e^{-\ell/2 } K_{i \sqrt{8E}}(e^{-\ell/2})
$  \cite{Yang:2018gdb}. 
Here, $E$ can be interpreted as the ADM mass observed on the asymptotic boundary and the conservation of this mass follows from the orthogonality relation
\es{orthogonality-relation}{
\int_\infty^\infty d\ell e^\ell \,\psi_{\text{Disk}, E}(\ell) \psi_{\text{Disk}, E'}^*(\ell) = \frac{\delta(E-E')}{\rho_{\text{Disk}}(E)}\,.
}
Above, $ \rho_{\text{Disk}}(E)$ is the density of states in JT gravity on the disk, given by $\rho_{\text{Disk}}(E) = \sinh(2\pi \sqrt{E})$\cite{Stanford:2019vob}. Using \eqref{rewriting-MCG}  we can thus rewrite the genus $g$ contribution,\footnote{The integral over $\omega$ in \eqref{rewriting-MCG} can thus be rewritten by separating the integral over the moduli of the two closed geodesics separating the trumpets from the bordered Riemann surfaces,  $ \int_{\frac{\cT_{h,1}}{\Mod(\cM_{h,1})}} \omega  = \int_0^\infty db\int_0^b d\tau \int_{\frac{\tilde \cT_{h,1}(b)}{\Mod(\tilde \cM_{h,1}(b))}} \omega$ $\bigg($and  $\int_{\frac{\cT_{g-1,2}}{\Mod(\cM_{g-1,2})}} \omega  = \int_0^\infty db_1\int_0^b d\tau_1 \int_0^\infty db_2\int_0^b d\tau_2 \int_{\frac{\tilde \cT_{g-1,2}(b_1, b_2)}{\Mod(\tilde \cM_{g-1,2}(b_1, b_2))}} \omega\bigg)$ where the $\tilde \,$ signifies that we are considering the moduli space of surfaces with one (or two) geodesic boundary of fixed length $b$ (or $b_1$ and $b_2$).}  
\es{two-point-function-vol-decomp}{
\< \Tr_\beta\le(\chi(x_1) \chi(x_2) \ri)\>_{\nsi,\,g} &\sim e^{S_0(1 - 2g)}\int  d\ell\, e^\ell \int db_1 b_1\, db_2 b_2 \,\psi_{\text{Trumpet},x}(\ell, b_1) \psi_{\text{Trumpet},\beta - x}(\ell, b_2) e^{-\Delta \ell} \\&\times \bigg[ \Vol_{g-1, 2}(b_1, b_2)  +  \sum_{h\geq 0}  \Vol_{g-h, 1}(b_1)  \Vol_{h,1}(b_2) \bigg]\,, 
}
where, in order to compactly write \eqref{two-point-function-vol-decomp}, and not have to consider cuts which involve the disk wavefunction separately, we have formally chosen $\Vol_{0, 1}(b)$ to be such that, after the integral over $b$, the disk wavefunction is recovered\footnote{This is possible since $ \Vol_{0, 1}(b)$ (or its Laplace transform) does not enter the topological recursion used to determine $ \Vol_{g, n}$. }
\be 
 \psi_\text{Disk,x}(\ell) \equiv \int_0^\infty dE \rho_\text{Disk}(E) \psi_\text{Disk, E}(\ell)e^{- x E} \equiv \int_0^\infty db b \,\psi_{\text{Trumpet},x}(\ell, b) \Vol_{0, 1}(b) \,.
\ee

This formula can be seen as a generalization of the result from \cite{Saad:2019lba} which computed the boundary-to-boundary two-point function up to genus one, up to arbitrary genus.  Inserting \eqref{wavefunction-trumpet} into \eqref{two-point-function-vol-decomp}, using the orthogonality relation \eqref{orthogonality-relation} and performing the integral over $\ell$ explicitly, we can more conveniently rewrite \eqref{two-point-function-vol-decomp} as 
\es{CandidateFinal}{
\<\Tr_\beta&\left( \chi(x_1) \chi(x_2) \right)\>_\nsi= \\ 
&= {4e^{-S_0}\ov Z(\beta)}\int_0^\infty dE_1 dE_2 \ \le\<\rho(E_1)\rho(E_2) \ri\>^\text{JT} e^{-E_1 (x_1-x_2)- E_2(\beta-x_1+x_2)} \, \sM_\De(E_1,E_2)\,,\\
\sM_\De(E_1,E_2)&={\abs{\Ga\le(\De+i (s_1+s_2) \ri)}^2\ov 2^{2\Delta+1}}\times{\abs{\Ga\le(\De+i (s_1-s_2)\ri)}^2\ov \Ga(2\De)}\,, \hspace{0.5cm} \text{with} \hspace{0.5cm}  s_{1,2} \equiv \sqrt{2 E_{1,2}}\,.
}
where $\le\<\rho(E_1)\rho(E_2) \ri\>^\text{JT}$ is the spectral two-point function in JT gravity. The leading disk contribution of \eqref{CandidateFinal} has been derived using a variety of techniques \cite{Yang:2018gdb, Mertens:2017mtv, Blommaert:2018oro, Kitaev:2018wpr, Iliesiu:2019xuh}, while the arbitrary genus generalization found in \eqref{CandidateFinal} was hinted at by studying two-point function insertions directly in the matrix integral dual to JT gravity \cite{Saad:2019pqd, Blommaert:2020seb}. In these works $\sM_\De(E_1, E_2)$ was identified with the squared matrix element $\abs{\chi}^2_{E_1, E_2} = \abs{\<E_1|\chi|E_2\>}^2$, evaluated in the energy basis.  The formula \eqref{CandidateFinal} can also be derived using a matrix model description of JT gravity coupled to a scalar field \cite{jafferismatrix}.\footnote{We thank Daniel Jafferis for explaining this to us.}

The spectral two-point function has been the subject of intense study in JT gravity and its matrix integral dual \cite{Saad:2019lba, Blommaert:2019wfy, Blommaert:2020seb} (and in more general matrix integrals: see \cite{mehta2004random,Guhr:1997ve,millerRMT} for useful reviews) and, as we shall review shortly, when $E_1 \to E_2$, it can be estimated even at the non-perturbative level.

A small consistency check of this formula is that   for $\De=0$, at leading order in the genus expansion we should get $\< \chi(x_1) \chi(x_2) \>_\nsi^{\De \to 0} = 1$, since for the disk there is a unique geodesic.\footnote{At higher genus in JT gravity there are an infinite number of geodesics and therefore the insertion of the sum over geodesics is not equivalent to the insertion of the identity operator.}   Indeed,
\es{De0}{
\sM_0(s_1,s_2)&={\abs{\Ga\le(i (s_1+s_2) \ri)}^2\ov 2}\times 2\pi \de(s_1-s_2)={\de(s_1-s_2) \ov 4 \,r\le({s_1+s_2\ov 2}\ri)}\,,\\
r(s)&\equiv {s\ov 2\pi^2}\sinh(2\pi s)\,,
}
where the latter coincides with the  leading order density of states in JT gravity, $\rho^\text{(JT)}_0(s)=e^{S_0}r(s)$, with $\rho^\text{(JT)}_0(s) ds = e^{S_0} r(s) ds = e^{S_0} \rho_\text{Disk}(E) dE$.\footnote{Even though $r(s)$ happens to be related to the JT density of states, it will turn out to have a more general role in our story and we will therefore denote it differently. }  Since to leading order in $e^{S_0}$ (LO) the spectral two-point function factorizes, we get
\es{F0}{
\< \chi(x_1) \chi(x_2) \>^{\,\Delta \to 0}_\nsi\vert_\text{LO}&={4e^{-S_0}\ov Z_0(\beta)}\int_0^\infty ds_1 ds_2 \ \rho_0^\text{(JT)}(s_1)\rho_0^\text{(JT)}(s_2)  e^{-\frac12\le(s_1^2 (x_1-x_2)+ s_2^2(\beta-x_1+x_2)\ri)}\, {\de(s_1-s_2) \ov 4\,r\le({s_1+s_2\ov 2}\ri)}\\
&={1\ov Z_0(\beta)}\int_0^\infty ds_1 \ \rho_0^\text{(JT)}(s_1)\exp\le[-\beta { s_1^2\ov 2}  \ri] = 1\,.
}

Finally, we should mention that the above analysis can be extended for more general theories of dilaton gravity, with a dilaton potential $V(\phi)$, such as those considered in \cite{Witten:2020wvy, Maxfield:2020ale}. As we shall explain shortly, this will be particularly useful in exhibiting the generality of our results for the time dependence of the ER bridge. For a particular class of dilaton potentials, these theories were shown to have a matrix integral dual, with a matrix potential that is modified from the one encountered in JT gravity. This results in a modified spectral two-point function $ \le\<\rho(E_1)\rho(E_2) \ri\>^{V(\phi)}$. In appendix \ref{app:general-dilaton-gravity-2-pt-function}, we show that, by restricting to the same class of dilaton potentials as those considered in \cite{Witten:2020wvy, Maxfield:2020ale}, the above result for the two-point function that only accounts for non-intersecting geodesics \eqref{CandidateFinal}, still holds in these more general class of dilaton gravity theories, provided we replace $\le\<\rho(E_1)\rho(E_2) \ri\>^{JT}$ with $ \le\<\rho(E_1)\rho(E_2) \ri\>^{V(\phi)}$. Since \eqref{CandidateFinal} is now correct for more generic matrix integrals, we will use $ \le\<\rho(E_1)\rho(E_2) \ri\>$ to denote the spectral two-point function in a generic one-cut matrix integral.\footnote{The attentive reader may have noticed that \eqref{F0} does not give $1$ for dilaton gravities beyond JT, since we get ${1\ov Z_0(\beta)}\int_0^\infty ds \ {\rho_0(s)^2\ov r(s)}\exp\le[-\beta { s^2\ov 2}  \ri] \neq 1$. The reason for this is that inserting defects to the disk geometry leads to multiple geodesics, hence this quantity gives an average of the number of geodesics weighted by the number of defect insertions. Nevertheless we continue to normalize $\expval{\ell(t)}$ with $Z(\beta)$, since it is the norm of the Hartle-Hawking state, and not  by $\< \chi(x_1) \chi(x_2) \>^{\,\Delta \to 0}_\nsi$.}

\section{The length of the Einstein-Rosen bridge in dilaton gravity}\label{sec:evaluation}

We are primarily interested in the time dependence of
\be
 \<\ell(t)\> \equiv \< Tr_\beta (\hat \ell)\> \equiv  -\lim_{\De\to0}{ \frac{  \partial\le\< \Tr_\beta\left(\chi\le(\frac{\beta}2 + i t \ri) \chi\le(0\ri) \right)\ri\>_\nsi }{\p \De}}
\ee
 Since $\le\< \chi(x_1) \chi(x_2) \ri\>_\nsi$ only depends on $\De$ through $\sM_\De$, we take the $\De$ derivative inside the integral in \eqref{CandidateFinal} and evaluate it on $\sM_\De$:
\es{Mder}{
\om&\equiv s_1-s_2\,,\qquad \bar{s}\equiv{s_1+s_2\ov 2}\,,\\
-\lim_{\De\to0}{\p \sM_\De \ov \p \De}&=\#\de(\om)-{1\ov 16\pi^2 r(\bar{s})r\le({\om\ov 2}\ri)}\,.
}
From the exercise in \eqref{F0} we learned that the $\de(\om)$ only gives time independent contributions, which we do not compute explicitly.\footnote{This constant piece actually diverges as $1/\De$. In the computation where we take $\De\to0$ first, this divergence shows up as $[\de(\om)]^2$, where one factor of $\de(\om)$ comes from \eqref{Mder}, while the other comes from a contact term in  $\le\<\rho(s_1)\rho(s_2) \ri\> $, see \eqref{SSSspectral}. A divergence was expected from the start: we had to work with a regularized sum \eqref{regsum} because higher genus surfaces have infinitely many geodesics of increasing length, and when the regulator $\De$ is removed a divergence arises.  } Finally, we obtain a compact expression for $\<\ell(t)\>$:
\es{Ct}{
\<\ell(t)\>=\text{const}-{e^{-S_0}\ov 4\pi^2 Z(\beta)} \int_0^\infty ds_1 ds_2 \ {\le\<\rho(s_1)\rho(s_2) \ri\>\ov r(\bar{s})r\le({\om\ov 2}\ri)} \exp\le[-\beta\le({\bar{s}^2\ov 2}+{\om^2\ov8}\ri) -{i\bar{s}\,\om t}\ri]\,.
}

For all one-cut matrix integrals, including those that describe the theories of dilaton gravity discussed in section \ref{sec:two-point-JT}, the form of $\le\<\rho(s_1)\rho(s_2) \ri\>$ is universal as $s_1 \to s_2$ ($E_1 \to E_2$).  Ref. \cite{Saad:2019lba} give the following approximate formula for the spectral correlations:
\es{SSSspectral}{
\le\<\rho(s_1)\rho(s_2) \ri\>\,=\,\,\underbrace{e^{2S_0}\hat{\rho}(s_1)\hat{\rho}(s_2)}_{\substack{\text{Semi-classical}\\ \text{disconnected piece}}}\,\,\, + \,\, \, \underbrace{e^{S_0}\hat{\rho}(\bar{s})\de(\om)}_\text{Contact term} \,\,  -\,\,   \underbrace{{1\ov  \pi^2 \om^2}\, \sin^2\le(e^{S_0}\pi \hat{\rho}(\bar{s})\,\om \ri)}_\text{Non-perturbative contribution}\,,
}
where we made all factors of $e^{S_0}$ manifest by defining $\rho_0(s)=e^{S_0}\hat{\rho}(s)$, we used that $E=s^2/2$ and $\rho$ transforms as a density, and in the last term that $\abs{s_1-s_2}\ll 1$. While there could be different non-perturbative completions for the matrix model associated to JT or other models of dilaton gravity, we emphasize that the formula \eqref{SSSspectral} is universal and common among all non-perturbative completions. Furthermore, given that \eqref{Ct} only depends on the spectral two-point function, one could in principle compute it even beyond random matrix theory in any quantum mechanical system and we will speculate further about the role that it has in chaotic systems in our discussion.

We can now perform the integrals in \eqref{Ct}. The second term in \eqref{SSSspectral} gives a $t$-independent contribution, and can be absorbed into the constant term. In fact this constant is divergent, since it is proportional to $\int d\om \ \de(\om)/\om^2$. As previously mentioned, a convenient renormalization prescription is to consider the quantity $\ell(t)-\ell(0)$, which is free of divergences. While this prescription will play a very important role in the computation of the variance in section~\ref{sec:variance-ER-bridge}, for ease of presentation here we will proceed without keeping track of the time-independent constant.

The contribution of the first term to the length of the ER bridge was considered in \cite{Yang:2018gdb}, where it was evaluated using a saddle point approximation and found to grow linearly in time. We proceed differently. First, we notice that because of the $ \exp\le[-{i\bar{s}\,\om t}\ri]$ factor for large $t$ the integral will be dominated by $\om\sim 1/t$, and implement this through the rescaling $\om=\hat{\om}/t$. 
 Using that $r\le({\om\ov 2}\ri)\approx \hat{\om}^2/(4\pi t^2)$, also defining $t=e^{S_0} \hat{t}$ and dropping the bar from $s$, we arrive at the expression valid at leading order in $t$ for both nontrivial terms in \eqref{SSSspectral}:
\es{Ct2}{
\<\ell(t)\>=\text{const}-{e^{2S_0} \hat{t}\ov \pi Z_0(\beta)} \int_0^\infty d{s} \int_{-\infty}^\infty &d\hat{\om} \ {\hat{\rho}^2({s})\ov r(s)\hat{\om}^2} e^{-\beta{{s}^2/ 2} -{i{s}\,\hat{\om}}} \\ &\times\bigg[\underbrace{1}_{\substack{\text{Semi-classical}\\ \text{disconnected}\\ \text{piece}}}-\,\,\underbrace{\le({\hat{t} \ov  \pi \hat{\rho}({s}) \hat{\om}}\ri)^2\, \sin^2\le({\pi \hat{\rho}({s})\,\hat{\om} \ov \hat{t}}\ri)}_\text{Non-perturbative contribution}\bigg] \,.
}
The $\hat{\om}$ integral can now be evaluated exactly. Remarkably, the semi-classical and non-perturbative contributions conspire such that 
the integrand is regular at $\hat{\om}=0$.
Because of this remarkable cancellation, for $\pi \frac{\hat \rho({s})}{s}<\hat{t}$ (i.e.~$\sinh(2\pi{s})<\pi\hat{t}$ in the case of JT gravity) the integral can be evaluated by closing the contour for the $\hat \omega$ integral using the arc at complex infinity in the upper-half plane. Since the arc does not contribute to the contour integral and the contour does not enclose any poles, we conclude that for  $\pi \frac{\hat \rho({s})}{s}<\hat{t}$ the $\hat \omega$ integral vanishes  {\it for any density of states $\hat{\rho}(s)$, hence any dilaton gravity model}.  

For  $\pi \frac{\hat \rho({s})}{s}>\hat{t}$ one can start by slightly deforming the contour of $\hat \omega$  in the lower half-plane to avoid  the point $\hat \omega=0$.\footnote{This does not have any effect on the value of the integral since the integrand is regular at $\hat \omega = 0$. } Expanding the sine-kernel into exponentials, one can now enclose the $\hat \omega$ integral in the upper or lower half plane in such a way that the arc at complex infinity never contributes; i.e.~for terms containing $e^{i \# \omega}$ with $\# < 0$ the contour is enclosed in the lower-half plane while for $\# > 0$ it is enclosed in the upper half plane. Each term can contain a pole at $\hat \omega = 0$ however only the integrals enclosing the upper-half plane receive contributions from the pole. Adding the contribution of the appropriate residues we find that 
\es{Ct3}{
\<\ell(t)\> &=\text{const}-{2\pi e^{2S_0} \ov 3 Z_0(\beta)} \int_{s_*(\hat{t})}^\infty d{s} \ {\hat{\rho}^3(s)\ov r(s)}\exp\le[-\beta{{s}^2/ 2}\ri]\le(1-X(\hat{t},s)\ri)^3 \,,\\
X(\hat{t},s)&\equiv{s\hat{t}\ov 2\pi \hat \rho(s)}\,, \qquad Z_0(\beta)={e^{S_0} \int_0^\infty ds \hat \rho(s) e^{-\beta s^2 \ov 2}}
}
where $s_*(\hat{t})$ is such that $\pi \frac{\hat \rho({s_*(\hat t)})}{s_*(\hat t)}=\hat{t}$. In the specific case of JT gravity, 
\be 
X(\hat{t},s)\equiv{\pi \hat{t}\ov \sinh(2\pi{s})}\,, \qquad s_*(\hat{t})={1\ov 2\pi}{\rm arcsinh}(\pi\hat{t})\,.
\ee
We can easily evaluate this expression for $\<\ell (t) \>$ numerically, see Fig. \ref{fig:1} for the case of JT gravity.

 We can also compute the asymptotics of $\< \ell(t)\>$ for small and large $\hat{t}$. To be concrete, we can consider the particular example of JT gravity even though the asymptotics for small and large times will qualitatively be the same.\footnote{The large asymptotics will be the same if $s_*(\hat t)\to \infty$ for $\hat t \to \infty$. This occurs if $\hat \rho(s)$  grows faster than linearly as $s \to \infty$.  } For large $\hat{t}$ the integral goes to zero, since the lower limit of the integral $s_*(\hat{t})\to\infty$ and the integrand decays as $\exp\le[-\beta{{s}^2/ 2}\ri]$ for large $s$.\footnote{ Note that $0\leq X(s)\leq1$ and hence $0\leq \le(1-X(s)\ri)^3\leq 1$, and $\hat{\rho}^2(s)\sim\exp(4\pi s)$.} The leading late time asymptotics is obtained by replacing the $\sinh$ functions with exponentials, and evaluating the integral explicitly to obtain:
\es{largeTime}{
\<\ell(t)\> &= \text{const}-e^{ S_0}A\,\frac{ \hat t^2 \exp\le[-\frac{\beta  \log ^2(2 \pi \hat t)}{8 \pi ^2}\ri]}{ \log ^2(2\pi \hat t)}+\dots\,,\\
A&=\frac{32\sqrt{2} \,\pi ^{9/2}  e^{-{2\pi^2/ \beta}}}{\beta ^{5/2} }
}
At small times, we evaluate the integral by changing variables to $\xi=\hat{t}X(s)$, expanding the resulting integrand for small $\hat{t}$ and evaluating the resulting integrals. We get that $\<\ell(t)\>= \text{const} -C_0+\hat C_1\hat{t}+O(\hat{t}^2)$. We then absorb $C_0$ in $\text{const} $ to obtain the final result:
\es{Asymp}{
\<\ell(t) - \ell(0)\> \approx 
\begin{cases}
\hat C_1\hat{t}+\dots \qquad& \hat t\ll1\,,\\
C_0-e^{ S_0}A\,\frac{ \hat t^2 \exp\le[-\frac{\beta  \log ^2(2 \pi \hat t)}{8 \pi ^2}\ri]}{ \log ^2(2\pi \hat t)}+\dots \qquad& \hat t\gg1\,,
\end{cases}
}
where the explicit formulas for the constants are:
\es{Consts}{
C_0&={ e^{S_0}\ov 12 \pi^2}\le[e^{{6\pi^2/ \beta}}\le(1+{16\pi^2\ov \beta}\ri)-e^{-{2\pi^2/ \beta}} \ri]\,,\\
\hat C_1&= e^{S_0}\left[{e^{-{2\pi^2/ \beta}}\ov \sqrt{\pi\beta/2}}+{{\rm erf}\le(\sqrt{2}\pi\ov\sqrt{\beta}\ri)\ov 2\pi}\le(1+{4\pi^2\ov \beta}\ri)\right]\,.
} 
In particular, the early-time linear growth is controlled by $\hat C_1$. At temperatures high compared to the boundary value of the dilaton (which we have set to one), we have $\hat C_1=2\pi e^{S_0}/\beta$, and hence $\<\ell(t)\>\approx \frac{2\pi}{\beta}t$, which is the expected linear growth of the extremal volume in the classical regime \cite{Harlow:2018tqv}.\footnote{By early-time here we mean $\hat t=e^{-S_0}t\ll1$, but $t \gg1$. Transient effects at $t\sim \beta$ have been dropped in \eqref{Ct2}.} At low temperatures, the Schwarzian theory becomes strongly coupled, and the early time growth coefficient receives loop corrections. In particular, at low temperatures, $\<\ell(t)\> \approx \frac{4}{\sqrt{2\pi \beta}}t$ at early times.

\section{A comparison to the spectral-form factor and the two-point function}
\label{sec:comparison-section}

It is insightful to compare $\<\ell(t)\>$ to other related quantities, the spectral form factor (SFF) and the two-point function. The approximation to the two-point function is given in \eqref{CandidateFinal}, while the spectral form factor (with inverse temperature $\beta/2$) is proportional to the same formula with the replacement $\sM_\De(s_1,s_2)\to 1$. Applying the same logic as for $\<\ell(t)\>$, we can expand the integrand for small $\om=\hat{\om}/t$, and obtain the formula as in \eqref{Ct2} 
\es{Ct2B}{
F(t)\propto\text{const}+ \int_0^\infty d{s} \int_{-\infty}^\infty d\hat{\om} \ &\hat{\rho}^2({s})\le[1-\le({\hat{t} \ov  \pi \hat{\rho}({s}) \hat{\om}}\ri)^2\, \sin^2\le({\pi \hat{\rho}({s})\,\hat{\om} \ov \hat{t}}\ri)\ri] e^{-\beta{{s}^2/ 2} -{i{s}\,\hat{\om}}}\,\, \times \\
&\times\,\begin{cases}
{1\ov r(s)\hat{\om}^2} \hspace{2.9cm} \text{for $\<\ell(t)\>$,}\\
1 \qquad\hspace{2.9cm} \text{for SFF$_{\beta/2}(t)$,}\\
{\abs{\Ga\le(\De+2is  \ri)}^2\ov 2^{2\Delta+1}}\times{\Ga\le(\De\ri)^2\ov \Ga(2\De)}\qquad \text{for $\expval{\chi_L(t) \chi_R(0)}$.}
\end{cases}
}
While these formulas are very similar, the presence of the $1/\hat{\om}^2$ singularity in the integrand of $\<\ell(t)\>$ leads to dramatic differences from the case of the SFF and the two-point function. For these latter two, the disconnected contribution represented by the $1$ in the brackets vanishes after performing the  $\hat{\om}$ integral.\footnote{The $\hat{\om}$ integral gives a $\de(s)$, which gets multiplied by an $s$ dependent function that vanishes at $s=0$.} Famously, it is the sine kernel that gives rise to both the ramp and plateau in the SFF and the two-point function. 
In contrast, for  $\<\ell(t)\>$, the initial linear growth was given by the disconnected contribution, and the role of the sine kernel was to {\it cancel this contribution} at very late times, giving rise to the plateau.\footnote{
One may wonder whether our approximation scheme can capture the slope of the SFF through the disconnected contribution to the spectral correlator. While, as explained above, at leading order in the $\hat{\om}/t$ expansion we got zero, at fourth-order in this expansion we encounter the Fourier integral $ \int d \hat{\om}\ \hat{\om}^4\exp\le[ -{i{s}\,\hat{\om}}\ri]=2\pi \de^{(4)}(s)$, which integrated against $\hat{\rho}^2({s})=s^4/\pi+\dots$ gives the first nonzero contribution. This is indeed the $1/t^3$ asymptotics of the slope of the SFF and higher orders of the $1/t$ expansion can be similarly recovered.
}

Curiously, after adding the disconnected and sine kernel contributions for all our quantities, we again get very similar $s$ integrals. Compare
\es{Ct3B}{
F(t)&\propto \text{const}-\int_{s_*(\hat{t})}^\infty d{s} \ \hat{\rho}^2(s)\exp\le[-\beta{{s}^2/ 2}\ri]\le(1-X(\hat{t},s)\ri) \times\begin{cases}
1 \qquad\hspace{2.9cm} \text{for SFF$_{\beta/2}(t)$,}\\
{\abs{\Ga\le(\De+2is  \ri)}^2\ov 2^{2\Delta+1}}\times{\Ga\le(\De\ri)^2\ov \Ga(2\De)}\qquad \text{for $\expval{\chi_L(t) \chi_R(0)}$,}
\end{cases}
}
to \eqref{Ct3}. The close connection between the behavior of the SFF and the two point function was emphasized in \cite{Saad:2019pqd,Cotler:2019egt}.

\begin{figure}[!t]
\centering
\includegraphics[width=0.65\textwidth]{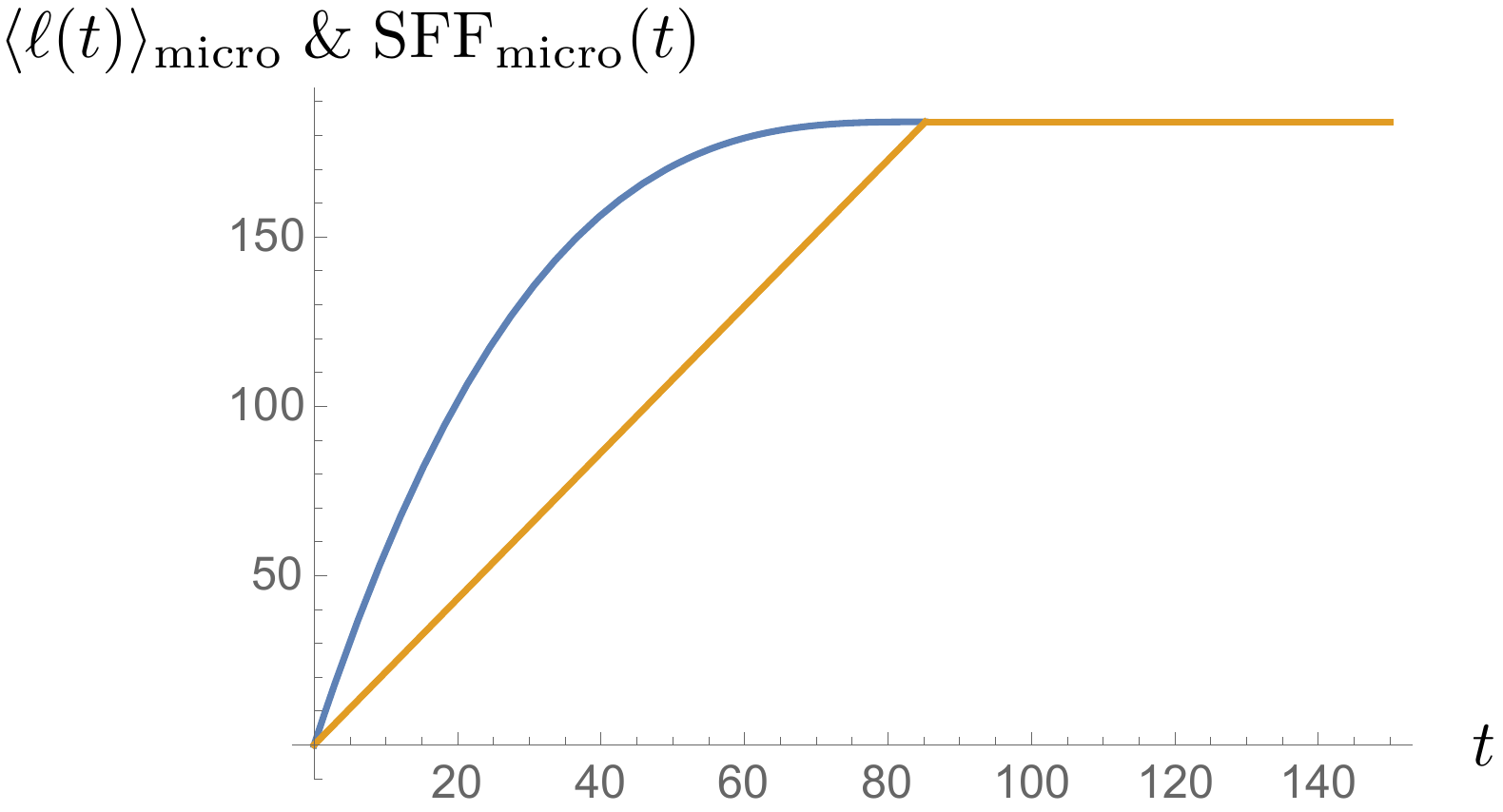} 
\caption{We show $\<\ell(t)\>_\text{micro}$ (blue) and $\text{SFF}_\text{micro}(t)$ (orange) for JT gravity with $s=1$ (or $E=1/2$). We are plotting the expressions in \eqref{micro2}, so we have dropped inessential prefactors. $\<\ell(t)\>$ is a cubic polynomial in time, while the SFF is its second time derivative (up to the addition of a constant). We have normalized $\<\ell(t)\>_\text{micro}$ and  $\text{SFF}_\text{micro}(t)$ in such a way that their plateau values are the same.  }
\label{fig:micro}
\end{figure}

We also note that we were surprised to find how slowly the integrals \eqref{Ct3} and \eqref{Ct3B} reach their asymptotic values for small $\beta$.  It is easy to repeat the analysis leading to  \eqref{largeTime}, and we  find $ \exp\le[-\frac{\beta  \log ^2(2 \pi \hat t)}{8 \pi ^2}\ri]$ asymptotics for all three quantities, as has been found before for the spectral form factor in the SYK model in \cite{Cotler:2016fpe}.

Finally, we analyze the microcanonical versions of these quantities. These are obtained by replacing the canonical density matrix $e^{-\beta H}$ with averaging over a microcanonical window $(E-\de,E+\de)$ with $\de\ll E$. 
Exploiting the identity
\es{Laplace}{
{1\ov 2\pi}\int_{c+i\mathbb{R}}d\beta\ e^{\beta E+\beta^2 \de^2}\, e^{-\beta {\cal E}}=2\pi\,{e^{-(E-{\cal E})^2/(4\de^2)}\ov \sqrt{4\pi \de^2}}\approx \de(E-{\cal E})\,,
}
we can convert each of our answers obtained in the canonical ensemble (or thermofield double state) by the integral transformation
\es{micro}{
F^\text{micro}_{(E,\de)}(t)=\int_{c+i\mathbb{R}}d\beta\ e^{\beta E+\beta^2 \de^2}  \, F_\beta(t)\,.
}
Then for all three quantities considered so far we get rid of the remaining $s$ integral:
\es{micro2}{
\expval{\ell(t)}_{(E,\de)}&\propto \text{const}-\Theta(s-s_*(\hat{t}))\,{\hat{\rho}^3(s)\ov s \, r(s)} \le(1-X(\hat{t},s)\ri)^3\,,\\
SFF_{(E,\de)}(t)&\propto\text{const}-\Theta(s-s_*(\hat{t}))\,{\hat{\rho}^2(s)\ov s } \le(1-X(\hat{t},s)\ri)\,,\\
\expval{\sO_L(t) \sO_R(0)}_{(E,\de)}&\propto\text{const}-\Theta(s-s_*(\hat{t}))\,{\hat{\rho}^2(s)\ov s } \le(1-X(\hat{t},s)\ri){\abs{\Ga\le(\De+2is  \ri)}^2\ov 2^{2\Delta+1}}\times{\Ga\le(\De\ri)^2\ov \Ga(2\De)}\,,
}
where $E=s^2/2$ and the extra $s$ in the denominator comes from the Jacobian when changing variables from $s$ to $E$ in the integral.\footnote{We note that  $SFF_{(E,\de)}(t)$ is sometimes denoted by $Y(E,\de,t)$ in the literature \cite{Saad:2018bqo,Saad:2019pqd}.} Both the ER bridge length and the two-sided correlator are best thought of as being calculated in a microcanonical thermofield double state (considered also in \cite{Stanford:2020wkf})
\es{microTFD}{
\ket{\text{TFD}_{(E,\de)}}=\int_{c+i\mathbb{R}}d\beta\ e^{\beta E+\beta^2 \de^2} \,\ket{\text{TFD}_\beta}\,.
}

Recall that $X(\hat{t},s)$ (defined in \eqref{Ct3}) is linear in time, hence before saturation $\expval{\ell(t)}_{(E,\de)}$ is a cubic and the other two quantities are linear functions of time. Up to unimportant $E$ (or $s$) dependent factors, the microcanonical versions the SFF and the two point function are the same functions of time, while the SFF is the second time derivative of  $\expval{\ell(t)}_{(E,\de)}$ up to the addition of a constant setting the size of the plateau.\footnote{This can already be anticipated from \eqref{Ct2B}, where in the $\hat{\om}$ integral the only difference between the ER bridge length and the SFF is the $1/\om^2$ factor in the former, roughly corresponds to double integration in time. The integral over $s$ however is nontrivial, so in the canonical ensemble ER bridge length and the SFF are not related in any simple way.} To emphasize this relationship, we plot these functions in figure \ref{fig:micro}. Nevertheless, this relationship does not imply that  $\expval{\ell(t)}_{(E,\de)}$ exhibits a plateau -- instead, the special cancellation discussed above is necessary for that to occur.

\section{Variance in the length of Einstein-Rosen bridge}
\label{sec:variance-ER-bridge}

Having computed the time dependence in the length of the ER bridge, we can use similar techniques to compute its variance. As usual, to compute the variance, we consider geometries with two asymptotic boundaries, with the geodesics ending on the two pairs of points located on the two boundaries. However, the definition of the variance suffers from several ambiguities: (i) in the connected geometries, it is unclear whether to sum over geodesics which connect the different asymptotic boundaries, and (ii) since we have restricted ourselves to geodesics that do not self-intersect, should we also only be considering geodesics that do not intersect with each other.  While we believe that the qualitative time-dependence of the variance is not heavily affected by these choices,  we will only consider geodesics that go between the same asymptotic boundaries, and we will restrict to non-intersecting geodesics.
\\\\

\begin{figure}[t!]
\centering
\includegraphics[width=0.9\textwidth]{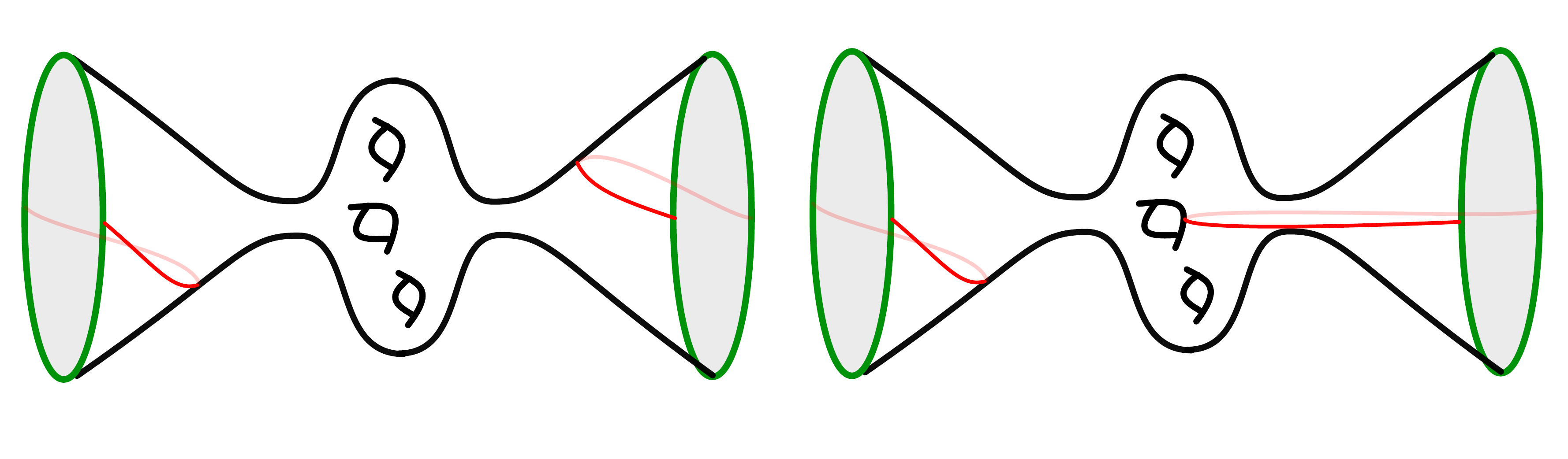} 
\includegraphics[width=0.9\textwidth]{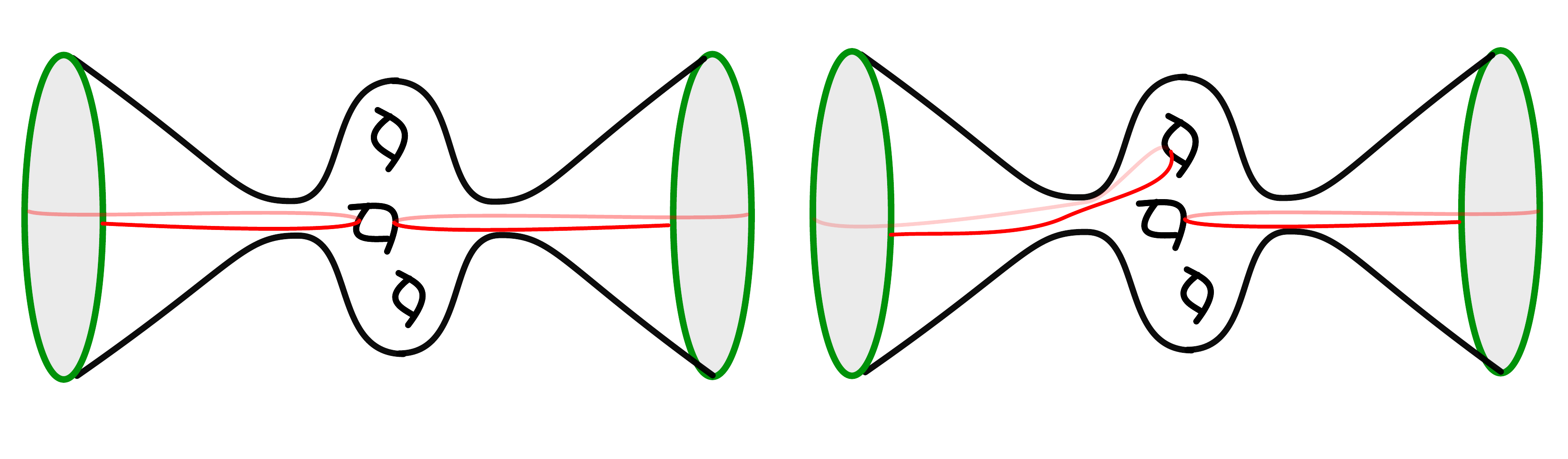} 
\caption{An example (with genus 3) of the type of hyperbolic surfaces which the path integral \eqref{general-formula-variance} sums over. The figures represent the cuts   $\cM_{g, 2} \to \cM_{h_1, 1}\oplus  \cM_{h_2, 1}\oplus \cM_{g-h_1 - h_2, 2}$ with $h_{1,2}\geq 0$ and $h_{1}+ h_2 \leq g$ (top left), $\cM_{g, 2} \to \cM_{h, 1} \oplus \cM_{g-h-1, 3}$ (top right), $\cM_{g, 2} \to \cM_{h, 2} \oplus \cM_{g-h-1, 2}$ (bottom left), and $\cM_{g, 2} \to \cM_{g-2,\, 4}$ (bottom right). }
\label{fig:example-at-genus-3-cylinder}
\end{figure}

\subsection{Geometric computation}
 To probe the variance for the length of the ER bridge we again use a modified boundary-to-boundary two-point function inserted on each asymptotic boundary, one for an operator $\chi$, with scaling dimension $\Delta$, and another for an operator $\chi'$, with scaling dimension $\Delta'$. We will denote the pair of  boundary-to-boundary two-point functions (with the assumptions about which geodesics contribute to the correlator as above) by $\<\Tr_\beta( \chi(x_1) \chi(x_2))\,\Tr_\beta( \chi(x_1') \chi(x_2') )\> _\nsi$. To compute the variance of the length, we will use 
\es{how-to-compute-variance}{
\sigma_{\<\ell(t)\>}^2 =\lim_{{\Delta \to 0, \Delta' \to 0}} \partial_{\Delta}  \partial_{\Delta'} \bigg[&  \< \Tr_\beta\le(\chi(x_1) \chi(x_2) \ri)\Tr_\beta\le(\chi'(x_1) \chi'(x_2) \ri)\>_\nsi \\ &- \,\,\< \Tr_\beta(\chi(x_1) \chi(x_2)) \>_\nsi \< \Tr_\beta(\chi'(x_1) \chi'(x_2) )\>_\nsi \bigg]\,.
}
The contribution coming from connected geometries is given by 
\es{general-formula-variance}{
\<\Tr_\beta &\le( \chi(x_1) \chi(x_2) \ri)\Tr_\beta\le(\chi'(x_1') \chi'(x_2') \ri)\>^{\,\text{connected}}_{\,\nsi} =  \sum_g e^{S_0(1-2g)} \\  &\times\int_{\frac{\cT_{g,2}}{\Mod(\cM_{g,2})}} \omega \int D{(\cW)}  D{(\cW')} e^{-I_{JT,\text{ bdy}}(\cW)} e^{-I_{JT,\text{ bdy}}(\cW')}  \sum_{\substack{\gamma \in \cG_{x1, x2}, \\\gamma' \in \cG_{x1', x2'},\\ \gamma \cap \gamma' = \emptyset}} e^{-\Delta \ell_{\gamma} - \Delta' \ell_{\gamma'}}
 } 
 where the integral over $\cW$ and $\cW'$ are the integrals over wiggles on the two asymptotic boundaries and the integral over $\omega$ is over the moduli space of hyperbolic manifolds with two asymptotic boundaries. Just like we discuss in section \ref{sec:two-point-JT} for the disconnected case or as we illustrate through an example in figure \ref{fig:example-at-genus-3-cylinder}, cutting along the pair of geodesics $\gamma$ and $\gamma'$ in the sum above, we can either obtain three or two disconnected manifolds or a single connected component.  In the first case, after cutting, $\cM_{g, 2} \to \cM_{h_1, 1}\oplus  \cM_{h_2, 1}\oplus \cM_{g-h_1 - h_2, 2}$ with $h_{1,2}\geq 0$ and $h_{1}+ h_2 \leq g$, in the second case, we can either obtain $\cM_{g, 2} \to \cM_{h, 1} \oplus \cM_{g-h-1, 3}$ or $\cM_{g, 2} \to \cM_{h, 2} \oplus \cM_{g-h-1, 2}$, and finally when obtaining a single connected component $\cM_{g, 2} \to \cM_{g-2,\, 4}$.\footnote{ $g=1$ requires special consideration, since in the third case we obtain two disconnected manifolds, thus reducing to the second case. }   Once again, the mapping class group acts by relating all pairs of boundary-to-boundary geodesics within a given cutting class as well as all closed geodesics (which do not intersect the boundary-to-boundary geodesics) on the surfaces,  $ \cM_{h_1, 1}\oplus  \cM_{h_2, 1}\oplus \cM_{g-h_1 - h_2, 2}$ in the first case, $\cM_{h, 1} \oplus \cM_{g-h-1, 3}$ in the second case, or  $\cM_{g-2,\, 4}$ in the third case. Therefore, as shown in appendix \ref{app:details-about-variance}, we present a generalization of \eqref{rewriting-MCG} for pairs of boundary-to-boundary geodesics. Putting together the results from that appendix, we find that  
\es{CandidateFinal-variance}{
\<\Tr_\beta &\le( \chi(x_1) \chi(x_2) \ri)\Tr_\beta\le(\chi'(x_1') \chi'(x_2') \ri)\>_{\,\nsi} ={4e^{-S_0}\ov \< Z(\beta) Z(\beta)\>}\int_0^\infty dE_1 \dots dE_4 \ \\ &\times  \le\<\rho(E_1)\rho(E_2) \rho(E_3) \rho(E_4) \ri\>  e^{-E_1 x_{12}- E_2(\beta-x_{12})-E_3 x_{12}'- E_4(\beta-x_{12}')} \, \sM_\De(E_1,E_2) \sM_{\De'}(E_3,E_4)\,,
}
where $\le\<\rho(E_1)\rho(E_2) \rho(E_3) \rho(E_4) \ri\>$ is the spectral four-point function in JT or dilaton gravity and $x_{ij} \equiv x_i - x_j$. By subtracting the product of two-point functions from \eqref{CandidateFinal-variance} we obtain the variance.  Of course, in the matrix integral interpretation of the theories of dilaton gravity, \eqref{CandidateFinal-variance} is the natural candidate for computing the square of \eqref{CandidateFinal} in an ensemble average. More generally, as discussed in appendix \ref{app:details-about-variance}, using the same geometric techniques we can compute any higher moments for $\<\Tr_\beta\le(\chi(x_1)\chi(x_2)\ri)\>_\nsi$.

\subsection{Evaluating the variance of the ER bridge length}
 Given that the dual of JT gravity is an ensemble, we want to quantify the size of the fluctuations in the volume. 
We therefore wish to evaluate $\langle \ell(t)^2\rangle$ using \eqref{CandidateFinal-variance}.
We take $\Delta\neq \Delta'$ in the regularization and therefore geodesics connecting distinct boundaries do not contribute. We take a $\Delta$ and a $\Delta'$ derivative of \eqref{CandidateFinal-variance} and set $\Delta=\Delta'=0$.
This results in the average of the square of complexity $\langle \ell(t)^2 \rangle$. 
To simplify the formulas and to incorporate the renormalization, we define the renormalized volume ``operator"
\beq
\label{eq:hatCdef}
\Tr_\beta (\hat \ell(t) - \hat \ell(0)) &= \int dE_1 dE_1 \rho(E_1)\rho(E_2) G(E_1,E_2)
\eeq
where the time dependence is in the function $G(E,E')$ and it reads as
\es{Ge1e2}{
G(E_1,E_2)=-{e^{-S_0}\ov 4\pi^2 Z(\beta)} \ {1 \ov r(\bar{s})r\le({\om\ov 2}\ri)} \exp\le[-\beta\le({\bar{s}^2\ov 2}+{\om^2\ov8}\ri) \ri] \le[\cos (\bar{s}\,\om t)-1\ri].
}
This kernel is obtained from formula \eqref{Ct}, after we have symmetrized the integrand in $\omega=s_1-s_2$ (the antisymmetric part does not contribute). Finally, we made a time independent subtraction so that the operator is $0$ at $t = 0$. This takes care of all the divergences, which we have seen to be time independent. Physically, it means that we set the volume of the ER bridge to be zero at zero time in all members of the matrix ensemble.

Now we want to evaluate $\< [\Tr_\beta (\hat \ell(t) - \hat \ell(0))]^2\>$. We need the four point density correlation in order to do this. In general, for a one matrix model, the $n$ eigenvalue marginals are given in terms of the Christoffel-Darboux kernel $K(E_1,E_2)$ (see \cite{millerRMT} for a concise pedagogical review):
\beq
\label{eq:Rs}
R_k(E_1,...,E_k)\equiv \frac{N!}{(N-k)!}\int dE_{k+1}...dE_N P(E_1,...,E_N) = \det K(E_i,E_j)|_{i,j=1,...,k},
\eeq
for instance $\langle\rho(E)\rangle=R_1(E)=K(E,E)$ and
\beq
\label{eq:2ptwithkernel}
\langle \rho(E_1)\rho(E_2)\rangle &= R_2(E_1,E_2)+\delta(E_1-E_2)R_1(E_1)
\\&=\langle\rho(E_1)\rangle\langle\rho(E_2)\rangle+\langle\rho(E_1)\rangle\delta(E_1-E_2)-K(E_1,E_2)^2.
\eeq
From this and \eqref{SSSspectral}, we read the kernel in the scaling regime $E_1 \to E_2$, relevant for the sine kernel:
\beq
\label{eq:Ksine}
K(s_1,s_2) \approx {1\ov  \pi \om}\, \sin\le(e^{S_0}\pi \hat{\rho}(\bar{s})\,\om \ri), \quad \quad \omega=s_1-s_2.
\eeq

\begin{figure}[!t]
\centering
\includegraphics[width=0.65\textwidth]{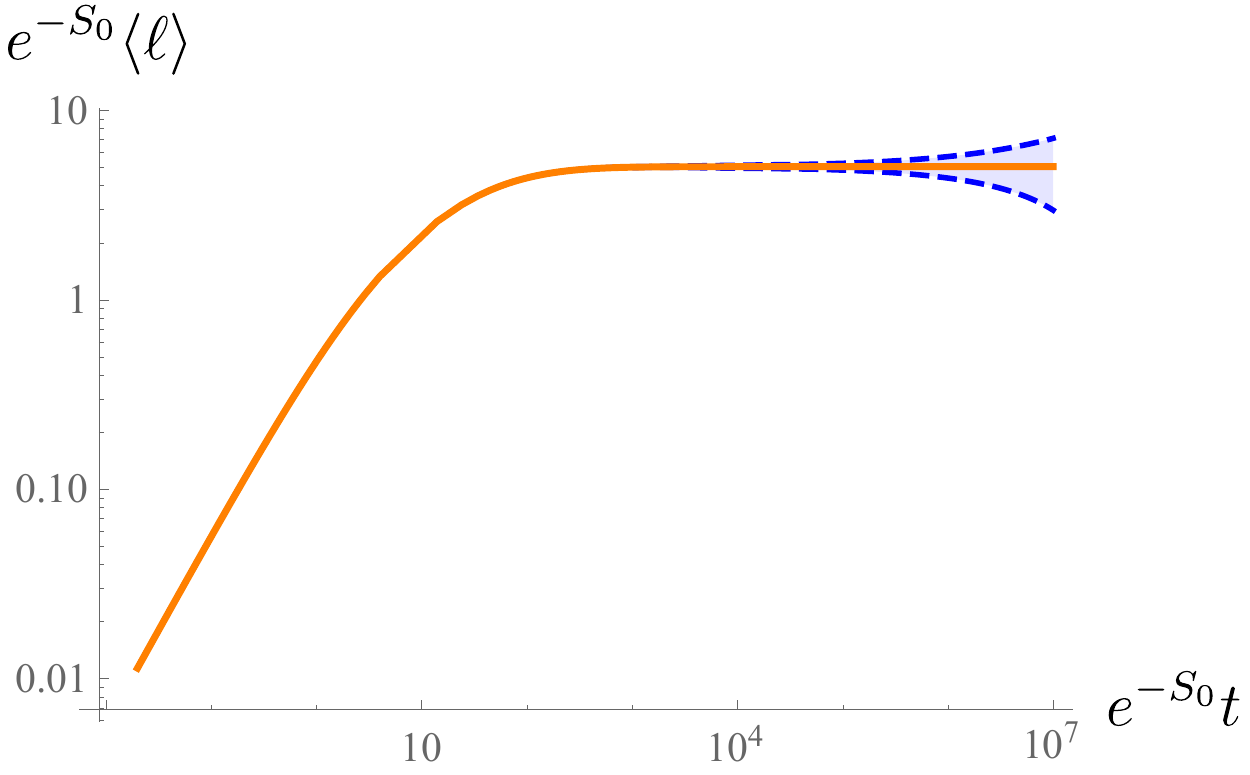} 
\caption{We show $\langle \ell \rangle$ and the band $\langle \ell \rangle \pm \sigma_{\ell}$, which at long times goes as $ e^{S_0}\pm \sqrt{t}$ giving the graph above on a log-log plot. We put $\beta=15$ and $S_0=10$ on this plot. The opening time of the band in $\hat t=e^{-S_0}t$ can be made arbitrarily late by increasing $S_0$.}
\label{fig:variance}
\end{figure}

The easiest way to deal with contact terms in density correlators is to write \eqref{eq:hatCdef} as a sum over the discrete spectrum in a single member of the matrix ensemble
\beq
\Tr_\beta (\hat \ell(t) - \hat \ell(0)) &= \sum_{i,j|E_i \neq E_j} G(E_i,E_j)
\eeq
 and directly integrate this expression against the joint distribution $P(E_1,...,E_N)$. Then we can use that the function $G$ is symmetric for exchange of its variables, and the $P$ and the $R_k$ are also totally symmetric. This way, we get that
\beq
\label{eq:varianceformula}
\sigma_\ell^2 &= \int dE_1...dE_4 [R_4(E_1,...,E_4)-R_2(E_1,E_2)R_2(E_3,E_4)]G(E_1,E_2)G(E_3,E4)\\
&+4 \int dE_1dE_2dE_3R_3(E_1,E_2,E_3)G(E_1,E_2)G(E_2,E_3) \\&+ 2 \int dE_1 dE_2 R_2(E_1,E_2)G(E_1,E_2)^2.
\eeq
where the second and third lines yield the contribution of contact terms in $\le\<\rho(s_1)\rho(s_2)\rho(s_3)\rho(s_4) \ri\>$, and we have already subtracted the disconnected contribution $\<\Tr_\beta (\hat \ell(t) - \hat \ell(0))\>^2$.
The dominant contribution to all these integrals come from the region where all energies are nearby. 
 Therefore, we can use \eqref{eq:Ksine} along with \eqref{eq:Rs}. Evaluating the integrals at large times results in\footnote{The limit of the last integral is not a misprint, we changed it from $s_*(t)$ to $s_*(2t)$ so that the integrand of the middle term simplifies.} 
 \beq
\sigma_\ell^2  &={e^{3 S_0}\ov \pi^2 Z(\beta)^2}\Big[ \int_{s_*(2 \hat t)}^\infty ds \frac{e^{-\beta s^2}}{r(s)^2} \frac{s^5 {\hat t}^5}{60 \pi}\\ & + \int_{s_*(\hat t)}^{s_*(2 \hat t)} ds \frac{e^{-\beta s^2}}{r(s)^2} \frac{(2\pi\hat \rho - \hat t s)^3(7 s^2 {\hat t}^2+2\pi s \hat t \hat \rho +8\pi^2 {\hat \rho}^2 )}{60 \pi}+\int_0^{s_*(2 \hat t)} ds \frac{e^{-\beta s^2}}{r(s)^2} \frac{2}{15}\pi^3 {\hat \rho}^4(5 s \hat t-7\pi \hat \rho) \Big],
\eeq
with $s_*(\hat t)$ as in \eqref{Ct3}.
The most important property of this expression is that for $\hat t =e^{-S_0}t=O(1)$, it is $O(e^{S_0})$. Since this is the square of the noise, we find that the signal to noise ratio is 
\be 
\frac{\sigma_\ell}\ell \propto e^{-\frac{S_0}{2}}\,.
\ee 
This is to be contrasted with the spectral form factor, for which this ratio is order one. At early times $t=O(1)$, the first integral dominates and the expression starts as $\propto e^{ S_0}{\hat t}^5=e^{-4S_0} t^5$, so at early times, the volume is reliably calculated by semi-classical gravity, as expected.\footnote{This is a non-trivial check of our prescription to calculate a non-perturbative volume by summing over all non-intersecting geodesics.} At late times, the last integral dominates, and we get a linear growth $\propto e^{S_0} \hat t$. Somewhat surprisingly, this growth never stops, and the signal to noise ratio becomes $O(1)$ at $\hat t\propto e^{S_0}$, or $t\propto e^{2S_0}$. We show this growing noise on the late parts of the plateau for JT gravity in figure \ref{fig:variance}.\footnote{Note that we can understand the source of this linear growth quite explicitly. Is is coming from the last term in \eqref{eq:varianceformula} involving $R_2$. At small $\omega$ relevant for late time, $R_2$, and its contribution to $\sigma_\ell^2$, read as
\beq
\frac{1}{\omega^4}\left(\rho^2-\frac{\sin^2 \pi \rho \omega}{\pi^2 \omega^2}\right) \approx \frac{\pi^2 \rho^4}{3\omega^2},\qquad \int d\omega  \frac{\pi^2 \rho^4}{3\omega^2} [\cos(s \omega  t)-1]^2=\frac{\pi^3\rho^4}{3}|s t|\,,
\eeq
respectively.} This might be because of the ambiguity in the definition of the variance since in our calculation we have not included the contribution of intersecting (but not self-intersecting) geodesics. Such intersecting configurations start existing for the genus one two-boundary geometry so we might expect them to affect the very late-time value of the variance.

The findings above present some tension with the interpretation of the volume as complexity. First, we expect complexity to have time independent fluctuations until the Poincar\' e recurrence time, which is $\propto \exp[\exp[S_0]]$. Second, for times $t\propto e^{S_0}$, we expect complexity to have O(1) fluctuations, instead of $e^{S_0/2}$ that we have found here.\footnote{We thank Adam Brown and Leonard Susskind for explaining this to us.} 
However, we should emphasize that the distribution of $\ell(t)$ seems to be highly non-Gaussian, since it is dominated by the region where all energies are nearby. This is opposed to the spectral form factor, where the four boundary correlator $\langle \rho \rho \rho \rho \rangle$ is well approximated by Wick contractions (two boundary wormhole contributions), indicating Gaussian noise. Because the noise is highly non-Gaussian, one should be careful in interpreting $\sigma_\ell$.
For example it could come from a distribution which has fixed noise near the origin, but has a growing bump for certain very large fluctuations. Resolving this requires one to study higher moments of $\ell$. This is in principle possible, but we leave it to future work. 
 In any case, to resolve this tension between the variance of the ER length and the expected variance of complexity, it would be interesting to explore all alternate definitions of the variance which we have described above.

\section{Discussion}
\label{sec:discussion}

\subsection{Summary}

To recapitulate, we have computed the length of the ER bridge at all times. At early times, this length grows linearly with time, as expected from a semi-classical analysis.\footnote{Even when accounting for the backreaction of the operator insertion on the metric \cite{Yang:2018gdb}.} At late times, it saturates at a time and value both proportional to $e^{S_0}$.  This saturation is due to a universal cancellation between the classical contribution and the non-perturbative corrections to the spectral two-point function. This type of saturation is distinct from the cancellation between the leading wormhole geometry and the non-perturbative corrections that is responsible for the plateau in the spectral form factor and in the probe matter two-point function. Also, as opposed to these latter quantities, we have found that for the length of the ER bridge, the ``noise'' on the plateau is much smaller than the signal, i.e.~$\sigma_\ell\ll \ell$ for $t \propto e^{S_0}$. However, we have also found that $\sigma_\ell$ grows forever and that it becomes the same size as the signal at $t\propto e^{2S_0}$. This raises some challenges in the interpretation of the volume as complexity. 
A possible resolution is that our geometric prescription to compute the variance breaks down at very late times. Indeed, in summing over pairs geodesics in two boundary geometries, we made a choice to only include geodesic pairs that do not intersect. It is possible that the contribution of these intersecting geodesics would change the picture at very late times and it would be interesting to investigate this in the future.

\subsection{A new spectral quantity}

Regarding the relation between the length of the ER bridge and complexity, one can take several different perspectives. The first perspective, which we present in this subsection, is that we can use the results in this paper to define the dual for the volume of the ER bridge in the boundary theory. Since our final formula solely depends on the spectral data in the matrix ensemble dual to dilaton gravity, one can imagine defining a similar quantity to \eqref{Ct} for any quantum system. Explicitly, for systems with discrete energy levels, one can define\footnote{Note that we replaced the function $r(s_1-s_2)$ in the denominator of \eqref{Ct} simply by $(E_1-E_2)^2$ since, at sufficiently late times both \eqref{Ct} and  \eqref{spectral-complexity}  are dominated by closely spaced energy pairs. Clearly, the spectral complexity can be defined for any ensemble of theories by averaging \eqref{spectral-complexity}.}
\es{spectral-complexity}{
\cC(t) \sim \sum_{E_1 \neq E_2}  {1 \ov  (E_1-E_2)^2} \exp\le[-\frac{\beta}2\le( E_1+E_2\ri)\ri]\le(\cos\le[ {(E_1-E_2) t}\ri] -1\ri)\,,} 
where we changed from the $s$ parametrization of energy to energy itself, which also changes the argument of the cosine compared to previous formulas (e.g. \eqref{Ge1e2}).
We shall call this quantity the \textit{spectral complexity} of the thermofield double state.  We believe that most chaotic systems have spectral statistics which closely follow that of random matrices, and hence, we expect $\cC(t) $ to have a similar behavior to the ER length computed in this paper. Once again, we expect it to start with a linear growth and, at a late time (order $e^{S}$), way past the thermalization time, asymptotes to a constant. Thus,  just like the complexity defined in \cite{Susskind:2014rva,Susskind:2014moa}, $\cC(t) $ grows past when conventional probes reach their thermal equilibrium.

While $\cC(t) $ is defined when preparing the $t=0$ state by the Hartle-Hawking prescription and is given by the thermofield double state in the boundary theory, one can also consider generalizing the formula for the length of the ER bridge for different states that are prepared by the Euclidean path integral with different boundary conditions. In turn, this would lead to a generalization of \eqref{spectral-complexity} for the spectral complexity of various states in the boundary theory. For instance, one generalization already discussed in this paper is described in section \ref{sec:comparison-section} where we have defined the length of the ER bridge in the microcanonical ensemble, i.e., when in gravity one fixes $\partial_n \phi$, instead of its canonical conjugate $\sqrt{g_{uu}}$ on the boundary of the dilaton gravity theory \cite{Goel:2020yxl}.

\subsection{Other Dyson ensembles}

\begin{figure}[!t]
\centering
\includegraphics[width=\textwidth]{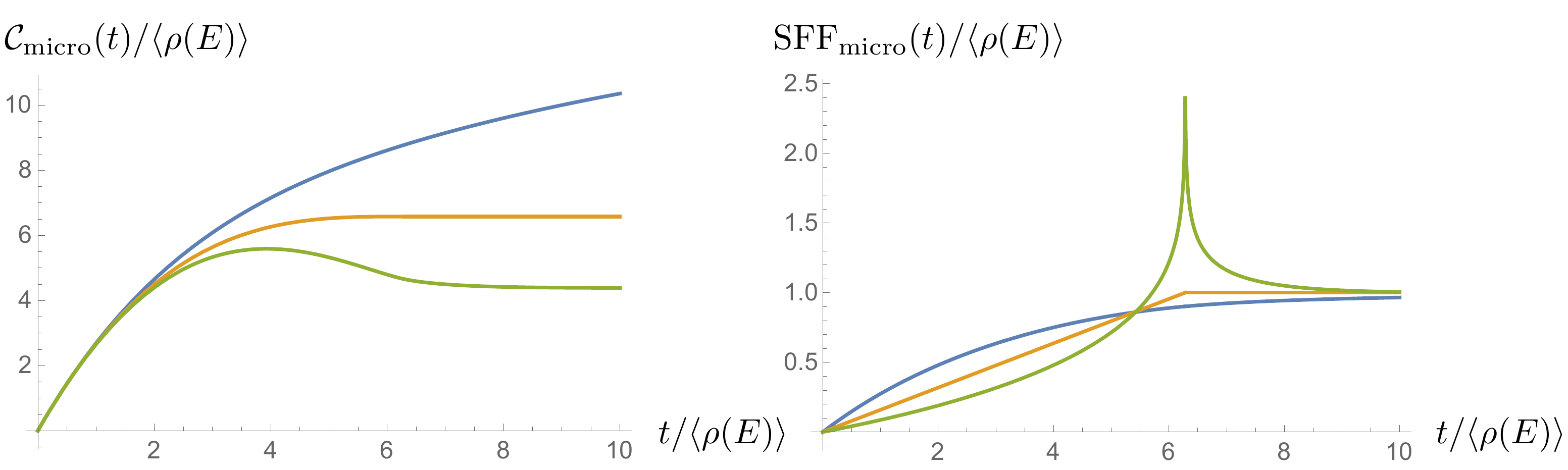} 
\caption{We show $\cC_\text{micro}(t)$ and $\text{SFF}_\text{micro}(t)$ for GOE (blue), GUE (orange) and GSE (green). }
\label{fig:Dyson}
\end{figure}

We can generalize the results in this paper to compute the spectral complexity
in any random matrix ensemble. The spectral correlators of dilaton gravity theories investigated above are universally governed by the Gaussian unitary ensemble (GUE) for small energy separations (or late times). It is an interesting question to ask, what is the behavior of $\mathcal C(t)$ in the other orthogonal (GOE) and symplectic (GSE) ensembles. While these ensembles are related to versions of JT gravity that includes a sum over unoriented surfaces, one encounters divergences for higher genus surfaces \cite{Stanford:2019vob}. At the level of the matrix integral this appears due to the double-scaling limit necessary in order to compute observables in two-dimensional gravity. Unless this divergence can somehow be resolved, this means that the computation below will not have the obvious interpretation of an ER bridge length in a theory of gravity. Nevertheless, it is interesting to understand the behavior of this quantity in matrix models. 

Towards that end, it is sufficient to work with the microcanonical version of the spectral complexity or ER bridge length discussed  around \eqref{micro2}. Up to inessential overall factors this is given by
\es{ellmicro}{
\cC_\text{micro}(t) =  - \int_{-\infty}^\infty &d\om \ {\expval{\rho(E+\om/2)\rho(E-\om/2)}\ov {\om}^2} \le[\cos({\om}t )-1\ri] \,.
}
Comparing with the microcanonical SFF
\es{SFFmicro}{
\text{SFF}_\text{micro}(t)= \int_{-\infty}^\infty &d\om \ {\expval{\rho(E+\om/2)\rho(E-\om/2)}} \le[\cos({\om}t )-1\ri] \,,
}
it is clear that 
\es{DerRel}{
\text{SFF}_\text{micro}(t)=\text{const}+{d^2\ov dt^2}\cC_\text{micro}(t)\,,
}
as was already discussed in section \ref{sec:comparison-section}. We give the density two point functions and the explicit formulas for both of these quantities in appendix \ref{app:RMT}, and plot them on figure \ref{fig:Dyson}. Curiously, $\cC_\text{micro}$ does not lead to a plateau in GOE, but grows logarithmically forever as $\expval{\rho(E)}\pi^2 \log(t)/3$. This can be anticipated from the fact that, while the SFF in GUE and GSE reaches a constant at finite $t$, in GOE its asymptotics is $\text{SFF}_\text{micro}(t)=\expval{\rho(E)}\le[1-\pi^2/(3t^2)+\dots\ri]$  and combining with the relation \eqref{DerRel}. Note however, that for all ensembles there is still a significant cancellation between the disconnected piece of the density two point function giving linear growth and the connected piece cancelling this at late times with the very slow asymptotic logarithmic growth remaining. Nevertheless, it is important to better understand the significance of this logarithmic growth in UV complete gravitational theories that can be related to individual members in the GOE ensemble.\footnote{It would also be interesting to study the behavior of the spectral complexity in supersymmetric ensembles \cite{altland1997nonstandard}, which should capture the behavior of the ER birdge in theories of supergravity \cite{Stanford:2019vob, Heydeman:2020hhw,Fan:2021wsb}. A first step in that direction was taken in \cite{Fan:2021wsb} who showed that the length of the ER birdge also grows linearly with time in $\mathcal N=1$ JT super-gravity.  }

\subsection{The quantity in SYK}

 \begin{figure}[!t]
\centering
\includegraphics[width=0.65\textwidth]{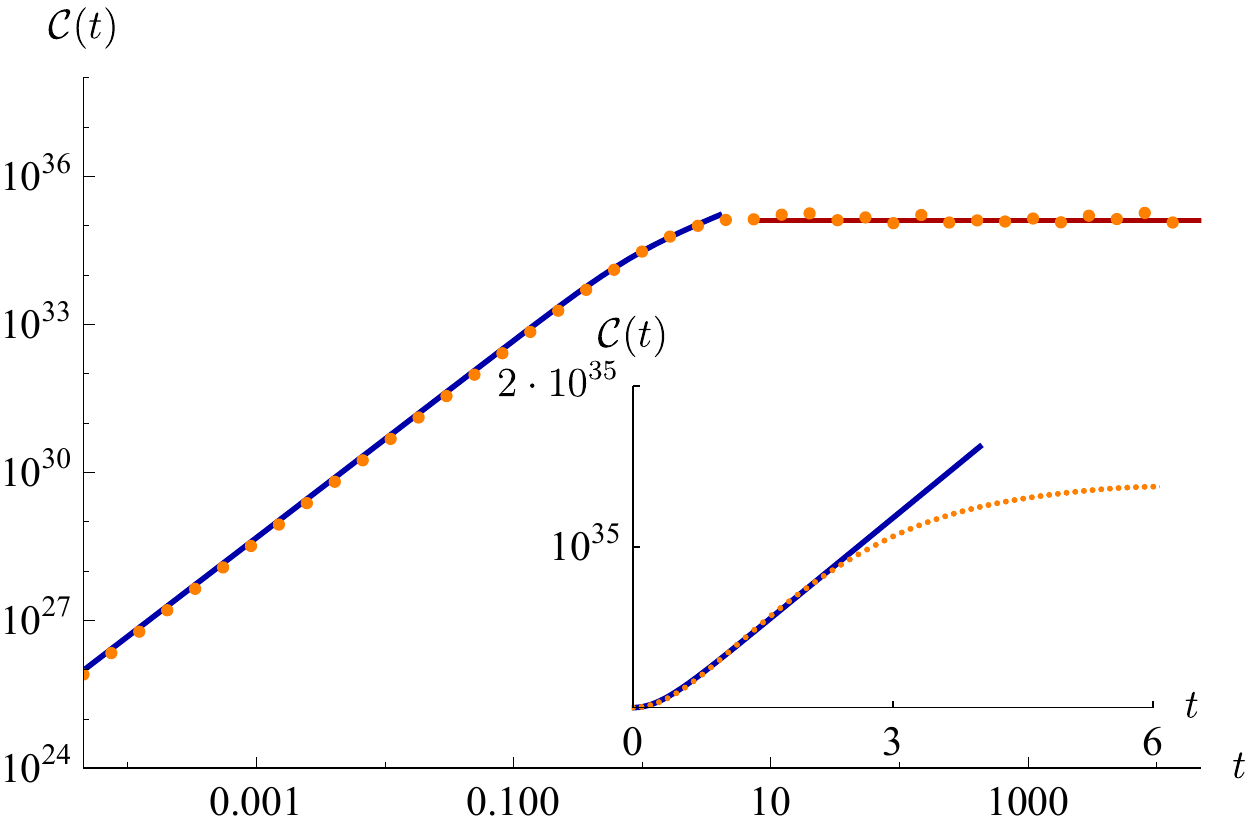} 
\caption{The spectral complexity in the SYK model with $N =18$, $q= 4$ and $\beta J=23$, averaged over $90$ members of the ensemble. Like in the random matrix theory example, the spectral complexity grows at early times as $\#\log \cosh(\# t/\beta)$ and plateaus at late times. The inset shows the linear scaled plot around $t = 0$ to better emphasize the validity of the logarithmic fit. }
\label{fig:SYK}
\end{figure}

Yet another concrete example in which we can compute the spectral complexity (this time numerically) that is different from the random matrix examples discussed throughout this paper, is the SYK model, a chaotic system with $q$-local interacting fermions \cite{sachdev1993gapless, kitaevTalks, Maldacena:2016hyu, Cotler:2016fpe}. In Figure \ref{fig:SYK} we show the spectral complexity for the $N=18$ and $q = 4$ SYK model at inverse temperature $\beta J = 23$. For this value of $N$ the corresponding ensemble is GUE \cite{Cotler:2016fpe}. Before thermalization time $\sim\beta$ the gravitational results suggest that the spectral complexity grows quadratically with time, after which it has a linear period of growth, followed by the plateau at later times. 

In accordance with expectations,  our results for $\cC(t)$ in SYK show an early time growth period followed by a plateau period. Due to the lower value of $N$ (due to computational constraints) as well as the higher value of $\beta$ (because we wish to see the plateau appear at a time which is computationally reachable) that we have simulated, the time difference between the thermalization time and plateau time is not parametrically large. Inspired by the semi-classical result for the ER bridge length we therefore fit the simulated values  of $\cC(t)$ at early times to $\ell(t) \propto \log \cosh(\# t/\beta)$.

\subsection{Relation to circuit complexity}

The second perspective on the relation between the ER bridge length and the complexity of the state is that $\cC(t) $ could be a good proxy for the circuit complexity of the thermofield double state in a variety of quantum systems.\footnote{Some previous efforts for defining complexity or a proxy for it in field theory with an eye towards holography include \cite{Miyaji:2015woj,Jefferson:2017sdb,Caputa:2017yrh,Chapman:2017rqy,Caputa:2018kdj,Belin:2018bpg,Erdmenger:2020sup,Kar:2021nbm}.}  If this is indeed the case, $\cC(t) $ would be a much easier quantity to compute since it only depends on the spectral statistics of the system. This brings up two challenges. The first is to check, at least in some systems where complexity is computable, whether the temperature dependence of  $\cC(t)$ matches that of complexity. The second challenge is that in the models of dilaton gravity discussed in this paper, there is no obvious notion of $k$-locality in the matrix integral duals. In the context of complexity, $k$-locality means that the Hamiltonian is a sum of terms that are products of at most $k$ operators that are considered to be simple (or local). In a theory without $k$-locality, one expects complexity to grow much faster than in $k$-local systems and saturate at its maximal value at much earlier times (perhaps not exponential in the entropy). Since $\cC(t)$ does not have such a behavior, it is difficult to see how it would be a proxy for complexity in such systems.\footnote{The definition of complexity in such matrix models is unclear. To define complexity, one needs to define a set of simple gates that can be used to produce the time-evolved state. In the ensemble average over random matrices, it is unclear whether to start with a unique set of gates for all members of the ensemble or whether to use a different set of gates for each member.  
} Nevertheless, it would be interesting to understand if in $k$-local chaotic  systems the spectral complexity indeed approximates the circuit complexity of the state.

\subsection{Back to gravity}

Finally, it would be interesting to understand the geometric origin of the plateau. In the case of the SFF \cite{Saad:2019lba} and the two-point function \cite{Saad:2019pqd}, a possible understanding of the plateau came from studying the effects of spacetime D-branes on the non-perturbative corrections to these observables in the matrix integral. In the case of the volume of the interior, we would need to understand the origin of the cancellation between the classical contribution and the non-perturbative contributions that come from D-branes.

It would also be important to understand if $\expval{\ell(t)}$ can be understood as the expectation value of a linear operator acting on some gravitational Hilbert space. This is a subtle question in gravity \cite{Jafferis:2017tiu,Marolf:2020xie}, and the circuit complexity in the dual boundary theory is certainly not a linear operator.

An  interesting extension of our work would be to consider single sided black holes that could be modeled by considering JT gravity with end of the world branes~\cite{Kourkoulou:2017zaj,Penington:2019kki,Gao:2021uro}. Finally, given that the models of two-dimensional gravity that we have studied are closely related to higher dimensional near-extremal black holes \cite{Sachdev:2015efa, Almheiri:2016fws,Sarosi:2017ykf, Nayak:2018qej, Moitra:2018jqs, Moitra:2019bub, Sachdev:2019bjn, Iliesiu:2020qvm}, we expect that our conclusions extend to the volume of such bridges for these more realistic black holes. However, it would be important to make this precise, since there are a plethora of geometries that the dimensionally reduced two-dimensional gravity theory does not take into account. 
Since black holes are chaotic systems, we believe that their spectral complexity defined in \eqref{spectral-complexity} will behave in the same way as analyzed in this paper.

\section*{Acknowledgements}

We thank Adam Brown, Daniel Jafferis, Douglas Stanford, Leonard Susskind, and Gustavo Joaq\'iun Turiaci for very useful discussions, Zhenbin Yang and Stephen Shenker  for thoughtful comments on the manuscript, as well as Jordan Cotler and Nicholas Hunter-Jones for correspondence.  MM is supported by the Simons Center for Geometry and Physics. LVI is supported by the Simons Collaboration on Ultra-Quantum Matter, a Simons Foundation Grant with No. 651440. The paper was finalized while LVI and MM were participating in the 
2021 Simons Summer Workshop.

\appendix

\section{The modified two-point function for dilaton potential $V(\phi)$}
\label{app:general-dilaton-gravity-2-pt-function}

To emphasize the universality of our JT gravity results, we can consider the two-point function as probe for the length of the ER bridge in more general models of dilaton gravity.\footnote{Such toy models of gravity also have two-sided black hole solutions with an ER bridge \cite{Witten:2020ert}.  } Our analysis will follow closely that of \cite{Turiaci:2020fjj, Witten:2020wvy}, combined with some of the basic building blocks discussed in section \ref{sec:two-point-JT}. We will start with models of dilaton gravity where the JT action is modified,\footnote{We follow the convention for $\alpha$ in \cite{Turiaci:2020fjj}, which is related to the convention for $\alpha$ in \cite{Witten:2020wvy} by $2\pi (1-\alpha) \to \alpha$.} 
 \es{dilaton-action-definition}{
 I_\text{dilaton} = I_{JT} - \lambda \int d^2 x \sqrt{g} e^{-2\pi(1-\alpha)\phi}\,,
 }
 and as in \cite{Turiaci:2020fjj, Witten:2020wvy}, for reasons that we shall review shortly, our techniques will solely apply when $0 < \alpha \leq 1/2$.\footnote{However, some of the results of \cite{Turiaci:2020fjj, Witten:2020wvy} were extended to $0< \alpha < 1$ by using the equivalence between the theory of dilaton gravity and  deformation of the $(2,p)$ minimal string theory \cite{Turiaci:2020fjj}. It would be interesting to see if the two-point function which we shall discuss below could also be understood as an observable in the deformed  $(2,p)$ minimal string theory. For the disk with a defect, this point was addressed in \cite{Mertens:2019tcm}.    }

Expanding the exponential, $\exp(-{I_\text{dilaton}})$, that appears in the path integral for the theory \eqref{dilaton-action-definition} in powers of $\lambda$, we have that 
\es{defect-insertion-potential}{
e^{-I_\text{dilaton}} \supset e^{-I_{JT}}\frac{\lambda^k}{k!} \int d^2 x_1 \sqrt{g} \dots \int d^2 x_k \sqrt{g} \left(e^{-2\pi(1-\alpha)\left[ \phi(x_1) + \dots \phi(x_k)\right]} \right)\,.
}
Integrating out $\phi(x)$ in the path integral of this dilaton gravity theory and pulling the integrals over coordinates out of the path integral,  we find that
\es{constraint-on-R}{
R(x) + 2 = 2\pi(1-\alpha)\sum_{i=1}^k \delta^2(x-x_i)\,.
}
Thus, the term in \eqref{defect-insertion-potential} can be understood as the insertion of $k$ conical-defects in a theory of JT gravity, where we then  integrate the defects over the entire manifold \cite{Mertens:2019tcm}. 

More generally,  we can deform JT gravity by adding the potential 
\es{decomposition-potential}{
U(\phi) =  \sum_{i=1}^r \lambda_i e^{-2\pi(1-\alpha_i) \phi}\,.
}
By expanding $e^{-I_\text{dilaton}}$ in powers of $\lambda_i$ and once again integrating-out the dilaton $\phi$, the term of order $k$ are equivalent to inserting $k$ defects in JT gravity whose deficit angles are among $\{\alpha_1, \dots, \alpha_r\}$. To simplify some of the formulas in this appendix, we will however focus on the theory \eqref{dilaton-action-definition} from which the generalization to any $U(\phi) $ in \eqref{decomposition-potential} (once again, with $ 0< \alpha_i < 1/2$) will immediately follow.

\begin{figure}[t!]
\centering
\includegraphics[width=0.9\textwidth]{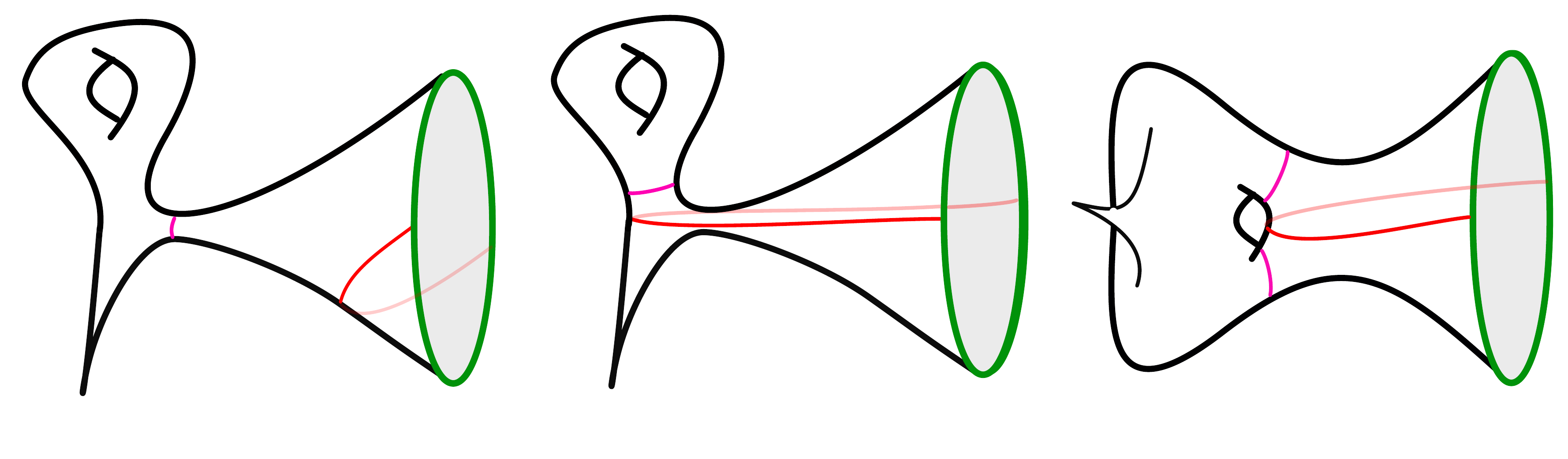} 
\caption{An example (with genus 1 and 1 defect) of the type of hyperbolic surfaces which the path integral with defect insertions sums over. Once again, while there are an infinite number of non-self-intersecting boundary-to-boundary geodesics on such surfaces, above we have drawn an example of all possible surface topology that can result by cutting a genus 1 and 1 defect surface along the geodesics. Once again, the purple curves represent the closed geodesics which we use to glue the trumpet wavefunctions in \eqref{two-point-function-vol-decomp} to different bordered Riemann surfaces. Compared to our computation in section \ref{sec:two-point-JT}, in the center figure we encounter a new ingredient as the decomposition of the manifold includes a ``defect wavefunction''. }
\label{fig:example-at-genus-1-defect}
\end{figure}

We can now study the insertion of the two-point function in the path integral of \eqref{dilaton-action-definition}. Once again expanding in powers of $\lambda$ we find that to evaluate the term $\sim \lambda^k$ we have to compute the two-point function in JT gravity in the presence of $k$ defects. This term is schematically given by 
\es{general-formula-w-defects}{
\< \Tr_\beta\le(\chi(x_1) \chi(x_2)\ri) \>_k = \frac{\lambda^k}{k!} \sum_g e^{S_0(1-2g)} \int_{\frac{\cT_{g,1,k}}{\Mod(\cM_{g,1,k})}} \omega \int D{(\cW)} e^{-I_{JT,\text{ bdy}}(\cW)}  \sum_{\gamma \in \cG_{x1, x2}^k} e^{-\Delta \ell_{\gamma}}\,,
 }
where the first sum is over surfaces of genus $g$, with $k$ defects,  the first integral is over the moduli space of such surfaces $\int_{{\cT_{g,n,k}}}\omega = \int_{{\cT_{g,n,k}}}\sum_{j=1}^{3g-3+n+k}db_j\wedge  d\tau_j$. We once again have to quotient this space by the mapping class group of each manifold, and, as before, the integral over  $D{(\cW)}$ gives the integral over the boundary wiggles in the presence of these defects. Finally, since we are interested in probing the volume of the ER bridge, we will only sum over geodesics that do not self-intersect, whose set we denote by $ \cG_{x1, x2}^k$. As in section \ref{sec:two-point-JT}, cutting along the boundary-to-boundary geodesic separates the surface into two disconnected components or yields a single connected component. In the former case, $\cM_{g,1, k} \to \cM_{h,1, p} \oplus \cM_{g-h, 1, k-p} $ where the geodesic can separate the set of defects to $p$ on one side (with $0 \leq p \leq k$), and $k-p$ on the other side. In the latter case, $\cM_{g,1, k} \to \cM_{g-1, 2, k}$ (see figure \ref{fig:example-at-genus-1-defect} for an example of both cases).\footnote{We do not expect the case, when the geodesic passes through a defect (or a set of defects) to significantly contribute to the path integral in \eqref{rewriting-MCG-defects}. This is because we are already summing over configurations in which the geodesics pass arbitrarily close to the defect and the geodesic length is continuous as the geodesic passes through the defect (since it is the length and not its derivative that appears in the action it is unimportant that the latter is discontinuous). } Once again, we have a trade-off between the integral over the moduli space and the sum over geodesics and can rewrite 
 \es{rewriting-MCG-defects}{
 \int_{\frac{\cT_{g,1,k}}{\Mod(\cM_{g,1,k})}} \omega \sum_{\gamma \in \cG_{x1, x2}} e^{-\Delta \ell_{\gamma}} &=e^{-\Delta \ell} \int_{\frac{\cT_{g-1,2, k}}{\Mod(\cM_{g-1,2,k})}} \omega\\  &+ \sum_{p\leq k}\sum_{h\geq 0} e^{-\Delta \ell}\int_{\frac{\cT_{h,1, p}}{\Mod(\cM_{h,1, p})}} \omega \int_{\frac{\cT_{g-h,1,k-p}}{\Mod(\cM_{g-h,1, k-p})}} \omega\,,
 }
where the geodesic of length $\ell$ is now on all the boundaries appearing for surfaces in  \eqref{rewriting-MCG-defects}. Having performed \eqref{rewriting-MCG-defects} we now want to break down all surfaces into further building blocks. For $0 < \alpha < 1/2$, surfaces of the type $\cM_{h, 1, p}$ (with $h\neq 0$ or $p>1$) or $\cM_{g-1, 2, k}$ appearing in \eqref{rewriting-MCG-defects} always have a closed geodesic (or two closed geodesics in the latter case) which separates all the defects and holes from the asymptotic boundary and the boundary-boundary geodesic.\footnote{This follows from a straightforward use of the Gauss-Bonnet theorem. Using the same theorem we can prove that this closed geodesic never intersects the boundary-to-boundary geodesics; if that is not the case, the resulting surface would include a bigon which would be disallowed in hyperbolic space \cite{farb2011primer}.   } In such a case, we can again separate our surface into trumpet wavefunctions and bordered Riemann surfaces of some genus $g$, with some number of boundaries $n$, and some number of defects $p$. The integral over the moduli space of such bordered Riemann surfaces  with defects gives the Weil-Peterson volume 
\es{volumes-Weil-Peterson-w-defects}{
\Vol_{g, n,k}(b_1,\,\dots,\,b_n; \alpha_1, \,\dots, \, \alpha_k) = \Vol_{g, n+k}(b_1,\,\dots,\,b_n, b_{n+1} =2\pi i \alpha_1, \,\dots, \, b_{n+k} = 2\pi i \alpha_k)\,,
}
which can be viewed as an analytic continuation of the usual Weil-Peterson volumes \cite{tan2006generalizations}.\footnote{ Hyperbolic surfaces have a natural $PSL(2, \mathbb R)$ connection. The holonomy $g$ around a cycle homotopic to a closed geodesic is given by hyperbolic $PSL(2, \mathbb R)$ elements. Similarly, one finds that the holonomy $g'$ around a cycle homotopic to a defect is given by elliptic $PSL(2, \mathbb R)$ elements. Since the Weil-Peterson volumes only depend on the conjugacy classes of the holonomies and the analytic continuation of the eigenvalues for an element in the hyperbolic class can yield an element in the elliptic class, \eqref{volumes-Weil-Peterson-w-defects} follows.  \label{holonomy-footnote}} There is however one exception to the decomposition above, when it involves a surface of the type  $\cM_{0, 1, 1}$; in  such a case, there is no closed geodesic separating the defect from the asymptotic boundary. We call this object, that has one defect and an asymptotic boundary glued to a boundary-to-boundary geodesic, a ``defect wavefunction''. The integral over the wiggles in \eqref{rewriting-MCG-defects} for such an object can be evaluated in precisely the same way as for the trumpet wavefunction. One finds,\footnote{As in footnote \ref{holonomy-footnote}, this follows from the fact that the result for the wavefunction can only depend on the conjugacy class of the holonomy around the closed geodesic of the trumpet or around the defect. Since the latter can be viewed as an analytic continuation of the former \eqref{wavefunction-defect} follows. }
\es{wavefunction-defect}{
\psi_{\text{Defect}, x}(\ell, \alpha) &=  \begin{tikzpicture}[baseline={([yshift=-.0ex]current bounding box.center)}, scale=0.6]
 \pgftext{\includegraphics[scale=0.5]{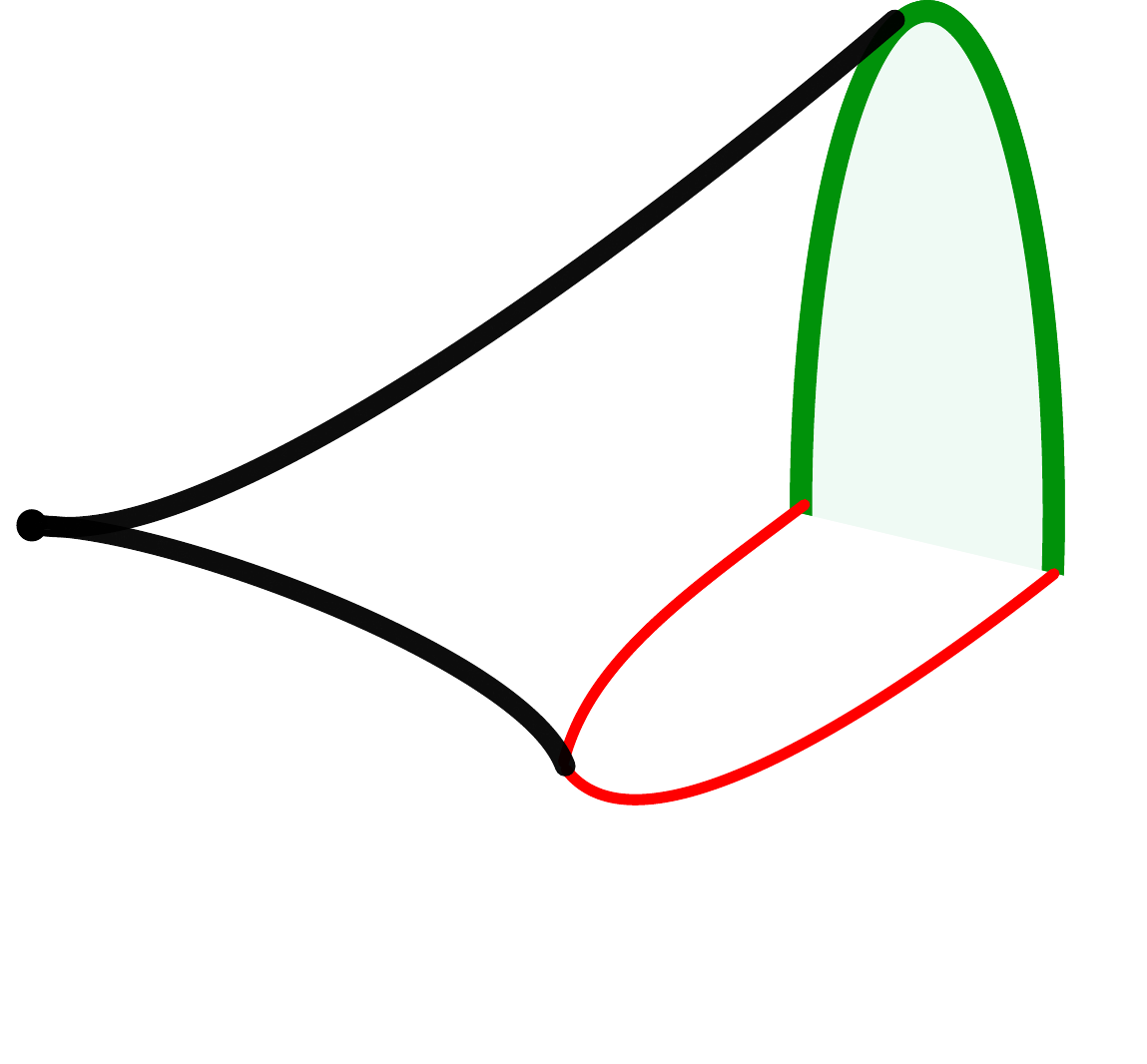}} at (0,0);
   \draw (-3.35,0) node  {$\alpha$}; 
    \draw (0.5,-1.7) node  {$\ell$}; 
  \end{tikzpicture} = \int_0^\infty dE \rho_{\text{Defect},\,b}(E) \psi_{\text{Disk, E}}(\ell) e^{-x E} \\ &= \psi_{\text{Trumpet}, x}(\ell, b= 2\pi i \alpha)\,,
}
with $\rho_{\text{Defect},\,\alpha}(E)  = \rho_{\text{Trumpet},\,b = 2\pi i \alpha}(E) $. Putting all these building blocks together we find that
\es{two-point-function-vol-decomp-w-defects}{
\< &\Tr_\beta (\chi(x_1) \chi(x_2)) \>_{g, k} \sim e^{S_0(1 - 2g)}\int  d\ell\, e^\ell \int db_1 b_1\, db_2 b_2 \,\psi_{\text{Trumpet},x}(\ell, b_1) \psi_{\text{Trumpet},\beta - x}(\ell, b_2) e^{-\Delta \ell} \\&\times  \lambda^k \bigg[ \frac{1}{k!}\Vol_{g-1, 2,k}(b_1, b_2;\underbrace{ \alpha, \,\dots, \, \alpha}_{k \text{ times}})\\  &\qquad \qquad + \sum_{p\leq k} \sum_{h\geq 0} \frac{1}{p!(k-p)!}  \Vol_{g-h, 1, p}(b_1; \underbrace{ \alpha,\dots, \alpha}_{p\text{ times}} )  \Vol_{h,1, p-k}(b_2; \underbrace{\pi i \alpha,\dots, 2\pi i \alpha}_{k-p\text{ times}} ) \bigg]\,,
}
where, in  order to not separately consider the decomposition into surfaces which involve a defect wavefunction, we will formally define $\Vol_{0, 1, 1}(b; \alpha)$ such that
\be
\int db b \Vol_{0, 1, 1}(b; \alpha)\psi_{\text{Trumpet},x}(\ell, b)=  \psi_{\text{Defect},x}(\ell, \alpha)\,.
\ee
\eqref{two-point-function-vol-decomp-w-defects} can then be compactly rewritten as
\es{putting-it-together-defect}{
\< \Tr_\beta (\chi(x_1)& \chi(x_2)) \> = \int_0^\infty dE_1 dE_2 \ \le\<\rho(E_1)\rho(E_2) \ri\>^{\text{dilaton}} e^{-E_1 (x_1-x_2)- E_2(\beta-x_1+x_2)} \, \sM_\De(E_1,E_2)\,,
}
where $\le\<\rho(E_1)\rho(E_2) \ri\>^{\text{dilaton}}$ is the spectral two-point function for the dilaton gravity theory \eqref{dilaton-action-definition}, which at order $\lambda^k$ has a genus $g$ contribution given by
\es{spectral-two-point-function-w-defects}{
 \le\<\rho(E_1)\rho(E_2) \ri\>^{\text{dilaton}}_{g, k} &\sim  \lambda^k \int db_1 b_1 db_2 b_2 \rho_{\text{Trumpet},\, b_1}(E_1) \rho_{\text{Trumpet},\, b_2}(E_2) \bigg[ \frac{1}{k!}\Vol_{g-1, 2,k}(b_1, b_2;\underbrace{ \alpha, \,\dots, \, \alpha}_{k \text{ times}})  \\ &+ \sum_{p\leq k} \sum_{h\geq 0} \frac{1}{p!(k-p)!}  \Vol_{g-h, 1, p}(b_1; \underbrace{ \alpha,\dots, \alpha}_{p\text{ times}} )  \Vol_{h,1, p-k}(b_2; \underbrace{\pi i \alpha,\dots, 2\pi i \alpha}_{k-p\text{ times}} ) \bigg]
}
Following precisely the same cutting and gluing algorithm, the formula \eqref{putting-it-together-defect} follows for any theory of dilaton gravity with a potential $U(\phi)$ as in \eqref{decomposition-potential}.

\section{Details about the geometric computation of the variance}
\label{app:details-about-variance}

In this section we will provide more details about the derivation of \eqref{CandidateFinal-variance}. The possible surfaces obtained from cutting along the two non-intersecting geodesics and the action of the mapping class group on geodesics on such surfaces in discussed in section \ref{sec:variance-ER-bridge}.  Just like in \eqref{rewriting-MCG}, we can rewrite 
 \es{rewriting-MCG-variance}{
 \int_{\frac{\cT_{g,2}}{\Mod(\cM_{g,2})}} \omega  &\sum_{\substack{\gamma \in \cG_{x1, x2}, \\\gamma' \in \cG_{x1', x2'},\\ \gamma \cap \gamma' = \emptyset}} e^{-\Delta \ell_{\gamma} -\Delta' \ell_{\gamma'}}=e^{-\Delta \ell -\Delta'\ell'} \int_{\frac{\cT_{g-2,4}}{\Mod(\cM_{g-2,4})}} \omega\\ &+ \sum_{h\geq 0} e^{-\Delta \ell - \Delta'\ell'} \left(\int_{\frac{\cT_{h,1}}{\Mod(\cM_{h,1})}} \omega \int_{\frac{\cT_{g-h-1,3}}{\Mod(\cM_{g-h-1,3})}} \omega +\int_{\frac{\cT_{h,2}}{\Mod(\cM_{h,2})}} \omega \int_{\frac{\cT_{g-h-1,2}}{\Mod(\cM_{g-h-1,2})}} \omega \right) \\ &+\sum_{h_1, h_2\geq 0} e^{-\Delta \ell-\Delta'\ell'}\int_{\frac{\cT_{h_1,1}}{\Mod(\cM_{h_1,1})}} \omega \int_{\frac{\cT_{h_2,1}}{\Mod(\cM_{h_2,1})}} \omega \int_{\frac{\cT_{g-h_1-h_2,2}}{\Mod(\cM_{g-h_1-h_2,2})}} \omega
 }
 where the geodesics $\ell$ and $\ell'$ are once again always on the boundaries of the manifolds resulting from the cut and where the sum over $h$ is again bounded such that disconnected surfaces of genus $h$ and $g-h$ are not counted twice and the sum over $h_{1, 2}$ are also bounded  such that all three disconnected surfaces within the same genus configuration are not counted multiple times. As before, we can once again then decompose our manifold into trumpet (or disk, when a component has $g=0$) wavefunctions and bordered Riemann surfaces. Peforming the integrals in \eqref{rewriting-MCG-variance} explicitly to obtain the volumes of these  bordered Riemann surfaces, we find that the genus $g$ contribution to $\< \Tr_\beta\le(\chi(x_1) \chi(x_2)\ri) \Tr_\beta\le( \chi(x_1') \chi(x_2') \ri)\>$, coming from connected geometries, is given by 
\es{two-point-function-vol-decomp-variance}{
\< \Tr_\beta &\le(\chi(x_1) \chi(x_2) \ri) \Tr_\beta \le( \chi(x_1') \chi(x_2') \ri)\>_{\text{connected, }g} \,\,\,\sim\,\,\,  e^{-2S_0 g}\int db_1 b_1\, db_2 b_2  db_3 b_3\, db_4 b_4 \, \\&\times \int  d\ell\, e^\ell\int  d\ell'\, e^{\ell'}  \psi_{\text{Trumpet},x}(\ell, b_1) \psi_{\text{Trumpet},\beta - x}(\ell, b_2)  \psi_{\text{Trumpet},x}(\ell', b_3) \psi_{\text{Trumpet},\beta - x}(\ell', b_4)\\&\times e^{-\Delta \ell - \Delta'\ell'}V_\text{connected}(b_1, b_2, b_3, b_4)
}
where, we define $V_\text{connected}(b_1, b_2, b_3, b_4)$ as,\footnote{We once again have to treat the components that have genus $g=0$ separately. }
\es{vol-generalized}{
V_\text{connected}&(b_1, b_2, b_3, b_4) = \Vol_{g-2, 4}(b_1, b_2, b_3, b_4)  \\ &+  \sum_{h\geq 0} \left[ \Vol_{h, 1}(b_1)  \Vol_{g-h,1}(b_2, b_3, b_4) + \Vol_{h, 1}(b_1)  \Vol_{g-h,1}(b_2, b_3, b_4)  \right] 
\\ &+  \sum_{h_1, h_2\geq 0} \left[ \Vol_{h_1, 1}(b_1) \Vol_{h_2}(b_2) \Vol_{g-h_1-h_2, 2}(b_3, b_4) \right]
}
Putting together the connected and disconnected contributions, writing the trumpet wavefunctions as in \eqref{wavefunction-trumpet}, and integrating over the $b$'s and $\ell$'s we can compactly rewrite the total two-boundary correlator $\< \Tr_\beta \le(\chi(x_1) \chi(x_2) \ri) \Tr_\beta \le( \chi(x_1') \chi(x_2') \ri)\>_\nsi$ as 
\es{CandidateFinal-variance-app}{
\< \Tr_\beta &\le(\chi(x_1) \chi(x_2) \ri) \Tr_\beta \le( \chi(x_1') \chi(x_2') \ri)\>_\nsi ={4e^{-S_0}\ov \< Z(\beta) Z(\beta)\>}\int_0^\infty dE_1 \dots dE_4 \ \le\<\rho(E_1)\dots  \rho(E_4) \ri\>  \\ &\times e^{-E_1 x_{12}- E_2(\beta-x_{12})-E_3 x_{12}'- E_4(\beta-x_{12}')} \, \sM_\De(E_1,E_2) \sM_{\De'}(E_3,E_4)\,.
}
More generally, using the same techniques to address the issues related to the mapping class group, we can compute even higher ``moments'' of the length of the ER bridge by using\footnote{The same ambiguities that exist for the variance, also exist for the higher moments and for the result \eqref{general-moment} we make the same choices as for the variance. }
\es{general-moment}{
\bigg\< \Tr_\beta &\le(\chi^{(1)}(x_1^{(1)}) \chi^{(1)}(x_2^{(1)}) \ri) \dots  \Tr_\beta\le( \chi^{(n)}(x_1^{(n)}) \chi^{(n)}(x_2^{(n)}) \ri) \bigg\>={4e^{-S_0}\ov \< \underbrace{Z(\beta) Z(\beta)\dots Z(\beta) \>}_{n \text{ copies}}} \\ &\times \int_0^\infty dE_1 \dots dE_{2n} \ \le\<\rho(E_1)\dots \rho(E_{2n}) \ri\>  \sM_{\De^{(1)}}(E_1,E_2)\dots \sM_{\De^{(n)}}(E_{2n-1},E_{2n})\\ &\times e^{-E_1 (x_1^{(1)}-x_2^{(1)})- E_2(\beta-x_1^{(1)}+x_2^{(1)})}  \dots   e^{-E_{2n-1} (x_1^{(n)}-x_2^{(n)})- E_{2n}(\beta-x_1^{(n)}+x_2^{(n)})} \, \,,
}
and by taking $\lim_{\Delta^{(1)},\,\dots\,,\Delta^{(n)} \to 0} \partial_{\Delta^{(1)}} \dots \partial_{\Delta^{(n)}}$ for the above expression.

\section{Explicit formulas in Dyson ensembles} \label{app:RMT}

 We largely follow the notation of \cite{Guhr:1997ve}. The density two point functions in Dyson ensembles are given by
 \es{rhorho}{
\expval{\rho\le(E+{\Om\ov 2\expval{\rho(E)}}\ri)\rho\le(E-{\Om\ov 2\expval{\rho(E)}}\ri)}&=\expval{\rho(E)}^2\le(1-Y(\Om)\ri)\,,\\
Y^\text{GOE}(\Om)&=s^2(\Om)+s'(\Om)\int_\Om^\infty d\nu\ s(\nu)\,,\\
Y^\text{GUE}(\Om)&=s^2(\Om)\,,\\
Y^\text{GSE}(\Om)&=s^2(2\Om)-s'(2 \Om)\int_0^\Om d\nu\ s(2\nu)\,,\\
s(\Om)&\equiv{\sin(\pi\Om)\ov \pi\Om}\,.
}
The SFF is given by formula \eqref{SFFmicro} and evaluates to 
\es{SFFapp}{
\text{SFF}^\text{GOE}_\text{micro}(2\pi \expval{\rho(E)} T)&=\expval{\rho(E)}\begin{cases}
2T-T\log\le(2T+1\ri)\qquad &(T\leq1)\,,\\
-T\log\le(2T+1\ov 2T-1\ri)\qquad &(T>1)\,.
\end{cases}\\
\text{SFF}^\text{GUE}_\text{micro}(2\pi \expval{\rho(E)} T)&=\expval{\rho(E)}\begin{cases}
T\qquad &(T\leq1)\,,\\
1\qquad &(T>1)\,.
\end{cases}\\
\text{SFF}^\text{GSE}_\text{micro}(2\pi \expval{\rho(E)} T)&=\expval{\rho(E)}\begin{cases}
\frac12T-\frac14T\log\abs{1-T}\qquad &(T\leq2)\,,\\
1\qquad &(T>2)\,.
\end{cases}
}
where we used a rescaled time variable $t=2\pi \expval{\rho(E)} T$. 
We can similarly compute $\cC(t)$ using \eqref{ellmicro} with the result 
\es{ellapp}{
\cC^\text{GOE}_\text{micro}(2\pi \expval{\rho(E)} T)&=\langle{\rho(E)}\rangle {\pi^2\ov 18}\begin{cases}
(30T-48T^2+34T^3)+(3+9T-12T^3)\log\le(2T+1\ri)\qquad &(T\leq1)\,,\\
(4+12T^2)+(3-9T+12T^3)\log\le(2T-1\ri)\\
+(3+9T-12T^3)\log\le(2T+1\ri)\qquad &(T>1)\,.
\end{cases}\\
\cC^\text{GUE}_\text{micro}(2\pi \expval{\rho(E)} T)&=\langle{\rho(E)}\rangle {2\pi^2\ov 3}\begin{cases}
3T-3T^2+T^3\qquad &(T\leq1)\,,\\
1\qquad &(T>1)\,.
\end{cases}\\
\cC^\text{GSE}_\text{micro}(2\pi \expval{\rho(E)} T)&=\langle{\rho(E)}\rangle  {4\pi^2\ov 9}\begin{cases}
{1\ov 16}\le[(60T-60T^2+17T^3)-(12-18T+6T^3)\log\abs{1-T}\ri]\qquad &(T\leq2)\,,\\
1\qquad &(T>2)\,.
\end{cases}
}
The relation \eqref{DerRel} is now easily verified.

\bibliographystyle{JHEP}
\bibliography{refs.bib}

\end{document}